\documentclass[aps,prd,reprint,superscriptaddress,floatfix,nofootinbib,showkeys,showpacs,preprintnumbers]{revtex4-2}
\usepackage{tabularx}
\usepackage{xspace}
\usepackage{listings}
\usepackage{lineno}
\usepackage{bm}		%
\usepackage{amsmath}	%
\usepackage{amssymb}	%
\usepackage[normalem]{ulem}

\usepackage[dvipsnames,svgnames]{xcolor} 
\usepackage{graphicx}
\usepackage{comment}

\usepackage{natbib}
\usepackage{hyperref}
\usepackage{amsmath}
\usepackage{enumitem}
\setlist{wide, labelwidth=!, labelindent=0pt}
\usepackage{multirow}
\hypersetup{    
  colorlinks      = true,
  linkcolor       = {black},
  linkbordercolor = {white},
  citecolor       = {black},
  citebordercolor = {white},
  urlcolor        = {blue},
  urlbordercolor  = {white},
}
\usepackage{soul} 
\usepackage{amsmath}
\usepackage{amssymb}
\usepackage{xspace}
\usepackage{xifthen}

\mathchardef\mhyphen="2D

\newlength{\dhatheight}

\newcommand{\bandvar}[2][]{%
  \ifthenelse{\isempty{#1}}{\var{#2}}{\var{#2\_#1}}%
}

\newcommand{\LCDM}{\ensuremath{\rm \Lambda CDM}\xspace}
\newcommand{\nuCDM}{\ensuremath{\rm \Lambda CDM}\xspace}

\newcommand{\var}[1]{\ensuremath{\texttt{\MakeUppercase{#1}}}\xspace}

\newcommand{\redmagic}{\rm redMaGiC}

\newcommand{\omegam}{\Omega_{\rm{m}}}
\newcommand{\sigmae}{\sigma_8}

\newcommand{\clustercomb}{CL+GC}
\newcommand{\allcomb}{CL+3$\times$2pt}
\newcommand{\ttt}{3$\times$2pt}
\newcommand{\hMpc}{h^{-1}\rm{Mpc}}

\providecommand\physrep{\ref@jnl{Phys.~Rep.}}%
\providecommand\apjs{\ref@jnl{ApJS}}%
\providecommand{\jcap}{\ref@jnl{JCAP}}%

\newcommand{\RNum}[1]{\uppercase\expandafter{\romannumeral #1\relax}}

\newcommand{\maglim}{\texttt{Maglim}}

\newcommand{\mcal}{\texttt{Metacalibration}}
\newcommand{\redmapper}{\texttt{redMaPPer}}

 \usepackage{lipsum}
\usepackage{aas_macros}

\usepackage{tikz}
\usepackage{calc}
\def\checkmark{\tikz\fill[scale=0.4](0,.35) -- (.25,0) -- (1,.7) -- (.25,.15) -- cycle;} 

\begin{document}
\preprint{DES-2025-0885}
\preprint{FERMILAB-PUB-25-0115-PPD}
\title{Dark Energy Survey Year 3 Results: Cosmological Constraints from Cluster Abundances, Weak Lensing, and Galaxy Clustering
 }

 \date{\today}
\author{T.~M.~C.~Abbott\textsuperscript{1}}
\author{M.~Aguena\textsuperscript{2}}
\author{A.~Alarcon\textsuperscript{3}}
\author{D.~Anbajagane\textsuperscript{4}}
\author{F.~Andrade-Oliveira\textsuperscript{5}}
\author{S.~Avila\textsuperscript{6}}
\author{D.~Bacon\textsuperscript{7}}
\author{M.~R.~Becker\textsuperscript{8}}
\author{S.~Bhargava\textsuperscript{9}}
\author{J.~Blazek\textsuperscript{10}}
\author{S.~Bocquet\textsuperscript{11}}
\author{D.~Brooks\textsuperscript{12}}
\author{A.~Carnero~Rosell\textsuperscript{13,2,14}}
\author{J.~Carretero\textsuperscript{15}}
\author{F.~J.~Castander\textsuperscript{16,3}}
\author{C.~Chang\textsuperscript{17,4}}
\author{A.~Choi\textsuperscript{18}}
\author{C.~Conselice\textsuperscript{19,20}}
\author{M.~Costanzi\textsuperscript{21,22,23}}
\author{M.~Crocce\textsuperscript{16,3}}
\author{L.~N.~da Costa\textsuperscript{2}}
\author{M.~E.~S.~Pereira\textsuperscript{24}}
\author{T.~M.~Davis\textsuperscript{25}}
\author{S.~Desai\textsuperscript{26}}
\author{H.~T.~Diehl\textsuperscript{27}}
\author{S.~Dodelson\textsuperscript{17,27,4}}
\author{P.~Doel\textsuperscript{12}}
\author{J.~Elvin-Poole\textsuperscript{28}}
\author{J.~Esteves\textsuperscript{29}}
\author{S.~Everett\textsuperscript{30}}
\author{A.~Farahi\textsuperscript{31,32}}
\author{A.~Fert\'e\textsuperscript{33}}
\author{B.~Flaugher\textsuperscript{27}}
\author{J.~Garc\'ia-Bellido\textsuperscript{34}}
\author{M.~Gatti\textsuperscript{4}}
\author{G.~Giannini\textsuperscript{15,4}}
\author{P.~Giles\textsuperscript{9}}
\author{S.~Grandis\textsuperscript{35}}
\author{D.~Gruen\textsuperscript{11}}
\author{R.~A.~Gruendl\textsuperscript{36,37}}
\author{G.~Gutierrez\textsuperscript{27}}
\author{I.~Harrison\textsuperscript{38}}
\author{S.~R.~Hinton\textsuperscript{25}}
\author{D.~L.~Hollowood\textsuperscript{39}}
\author{K.~Honscheid\textsuperscript{40,41}}
\author{N.~Jeffrey\textsuperscript{12}}
\author{T.~Jeltema\textsuperscript{39}}
\author{E.~Krause\textsuperscript{42}}
\author{O.~Lahav\textsuperscript{12}}
\author{S.~Lee\textsuperscript{43}}
\author{C.~Lidman\textsuperscript{44,45}}
\author{M.~Lima\textsuperscript{46,2}}
\author{H.~Lin\textsuperscript{27}}
\author{J.~J.~Mohr\textsuperscript{47,48}}
\author{J.~L.~Marshall\textsuperscript{49}}
\author{J.~McCullough\textsuperscript{50,51,33,11}}
\author{J. Mena-Fern\textsuperscript{52}}
\author{R.~Miquel\textsuperscript{53,15}}
\author{J.~Muir\textsuperscript{54,55}}
\author{J.~Myles\textsuperscript{50}}
\author{R.~L.~C.~Ogando\textsuperscript{56}}
\author{A.~Palmese\textsuperscript{57}}
\author{M.~Paterno\textsuperscript{27}}
\author{A.~A.~Plazas~Malag\'on\textsuperscript{51,33}}
\author{A.~Porredon\textsuperscript{6,58}}
\author{J.~Prat\textsuperscript{17,59}}
\author{A.~K.~Romer\textsuperscript{9}}
\author{A.~Roodman\textsuperscript{51,33}}
\author{E.~Rozo\textsuperscript{60}}
\author{E.~S.~Rykoff\textsuperscript{51,33}}
\author{E.~Sanchez\textsuperscript{6}}
\author{D.~Sanchez Cid\textsuperscript{6,5}}
\author{I.~Sevilla-Noarbe\textsuperscript{6}}
\author{M.~Smith\textsuperscript{61}}
\author{E.~Suchyta\textsuperscript{62}}
\author{G.~Tarle\textsuperscript{29}}
\author{D.~Thomas\textsuperscript{7}}
\author{Chun-Hao To\textsuperscript{17}}
\author{M.~A.~Troxel\textsuperscript{63}}
\author{V.~Vikram\textsuperscript{8}}
\author{A.~R.~Walker\textsuperscript{1}}
\author{David H. Weinberg\textsuperscript{40,64}}
\author{N.~Weaverdyck\textsuperscript{65,66}}
\author{R.~H.~Wechsler\textsuperscript{67,51,33}}
\author{J.~Weller\textsuperscript{48,68}}
\author{H.-Y.~Wu\textsuperscript{69}}
\author{M.~Yamamoto\textsuperscript{50,63}}
\author{B.~Yanny\textsuperscript{27}}
\author{Y.~Zhang\textsuperscript{1}}
\author{C.~Zhou\textsuperscript{63,39}}

\collaboration{DES Collaboration}
\date{Affiliations at the end of the paper. Contact author: \url{des-publication-queries@fnal.gov}}
\begin{abstract}
Galaxy clusters provide a unique probe of the late-time cosmic structure and serve as a powerful independent test of the \LCDM model. This work presents the first set of cosmological constraints derived with $\sim16,000$ optically selected redMaPPer clusters across nearly $5,000\ \rm{deg}^2$ using DES Year 3 data sets. Our analysis leverages a consistent modeling framework for galaxy cluster cosmology and DES-Y3 joint analyses of galaxy clustering and weak lensing (\ttt{}), ensuring direct comparability with the DES-Y3 \ttt{} analysis. We obtain constraints of $S_8 = 0.864 \pm 0.035$ and $\omegam = 0.265^{+0.019}_{-0.031}$ from the cluster-based data vector. We find that cluster constraints and \ttt{} constraints are consistent under the \LCDM{} model with a Posterior Predictive Distribution (PPD) value of $0.53$. The consistency between clusters and \ttt{} provides a stringent test of \LCDM{} across different mass and spatial scales. Jointly analyzing clusters with \ttt{} further improves cosmological constraints, yielding $S_8 = 0.811^{+0.022}_{-0.020}$ and $\omegam = 0.294^{+0.022}_{-0.033}$, a $24\%$ improvement in the $\omegam-S_8$ figure-of-merit over \ttt{} alone.  Moreover, we find no significant deviation from the Planck CMB constraints with a probability to exceed (PTE) value of $0.6$, significantly reducing previous $S_8$ tension claims. Finally, combining DES \ttt{}, DES clusters, and Planck CMB places an upper limit on the sum of neutrino masses of $\sum m_\nu < 0.26$ eV at 95\% confidence under the \LCDM{} model. These results establish optically selected clusters as a key cosmological probe and pave the way for cluster-based analyses in upcoming Stage-IV surveys such as LSST, Euclid, and Roman.

\keywords{Cosmology, Cosmological parameters, Galaxy cluster counts,  Large-scale structure of the universe}
\pacs{98.80.-k, 98.80.Es, 98.65.-r}
\end{abstract}
\maketitle 
\maketitle 
\section{Introduction}

The standard \LCDM model of cosmology has been successful in explaining a wide range of observational results \cite{Riess98, Perlmutter99, Suzuki12, WMAP9, Betoule14, BOSS17, Scolnic18, Planck18_params, DESY1KP, Y3kp, 4x2pt2, DES_SN, DESI_BAO, DESI24_FS} (see \cite{Frieman08, Weinberg_2013, Huterer15, Turner22, Huterer23} for reviews). However, recent evidence has started to hint the limitation of $\Lambda$CDM. Specifically, tension has emerged in the measurements of the $S_8$ parameter, the amplitude of the matter density fluctuations, defined as $S_8 = \sigma_8\sqrt{\omegam/0.3}$. The measurements of $S_8$ derived from the cosmic microwave background (CMB) \cite{Planck18_params}, when converted to today's values, tend to be higher than the late-universe values directly measured from large-scale structure 
\cite{Hikage19, Hamana20,  heymans2020kids1000, Asgari21, SeccoSamuroff22, AmonGruen22, Dalal23, Li23}, see \cite{DiValentino21, Perivolaropoulos22, Abdalla22} for reviews. Possible explanations of the tension range from unexpectedly strong baryonic feedback to beyond \LCDM physics \cite{AmonEfstathiou22, Preston23, Rogers23}. Another hint arises from measurements of the Hubble constant using local distance ladders, which yield a higher value than the one inferred from the CMB \citep[e.g.,][]{2020ApJ...891L...1P}. Moreover, recent combinations of Type-Ia supernovae (SN), Baryonic Acoustic Oscillations (BAO), and CMB show hints that the dark energy, which drives the universe's accelerated expansion, might not be a cosmological constant ($\Lambda$) \citep{2024arXiv240403002D, SNBAO}. 

To confirm or resolve the tension and to seek new physics beyond $\Lambda$CDM, we must examine the universe from multiple perspectives. Galaxy clusters (CL), galaxy clustering (GC), and weak gravitational lensing (WL) are each sensitive to different aspects of the late-time cosmic structure. Consistently analyzing, comparing, and combining insights from all these probes forms the foundation of multiprobe cosmological analysis and is a key goal of the Dark Energy Survey \citep{Y6BAO, Y3kp, Y5SN, SNBAO}. Among these probes, galaxy clusters are megaparsec scale structures that probe the late-time cosmic structure and have long been recognized as a powerful cosmological probe \cite{Haiman01, Holder01, LimaHu04, Allen11, Weinberg_2013}.
Cosmological analyses have been conducted using clusters identified in X-ray \cite{Vikhlinin09, wtg4,  2022A&A...663A...3G, Chiu23, erasscluster}, millimeter \cite{Bocquet19, SPTcosmo, SPTand3x2pt}, and optical surveys \cite{Rozo10, Costanzi19SDSS, To2021, KIDScluster, Sunayama2023, Fumagalli24}.

Wide-field imaging surveys, such as the Dark Energy Survey (DES), the Hyper Suprime Cam (HSC), and the Kilo-Degree Survey (KiDS), simultaneously provide a large sample of optically identified clusters and the gravitational lensing signal for cluster mass calibration.  
Forecasts have shown that the clusters have a statistical power comparable to that of combined CMB+SN+BAO+WL in Stage-III and Stage-IV experiments \cite{Weinberg_2013}.
However, despite its superb statistical power, optical cluster samples face unique challenges in systematic uncertainties.
Previous analyses have revealed that clusters selected by optical richness tend to suffer a selection bias in lensing \cite{DES_cluster_cosmology}. Specifically, without taking the selection bias in lensing into account, the cosmological constraints of $\sigma_8$ and $\omegam$ in the DES-Y1 small-scale analysis \cite{DES_cluster_cosmology} can be biased by more than $2\sigma$. Recent analyses have treated the cluster selection bias using either analytic or simulation-based approaches \cite{4x2pt1_duplicated, Sunayama2023, Salcedo23}.

\begin{figure}
    \centering
    \includegraphics[width=0.5\textwidth]{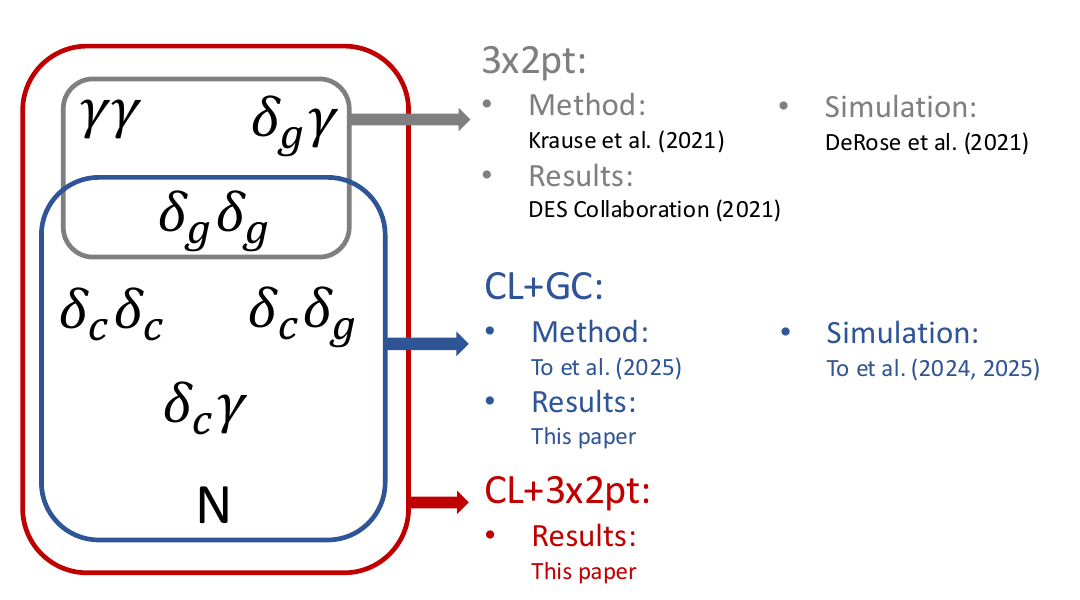}
    \caption{Summary of the different components in this analysis. The data in this paper consist of cluster abundances (N) and six two-point correlation functions derived from galaxy density ($\delta_g$), weak gravitational lensing shear ($\gamma$), and cluster density ($\delta_c$). The correlation functions include cosmic shear ($\gamma \gamma$), galaxy--galaxy lensing ($\delta_g \gamma$), galaxy clustering ($\delta_g \delta_g$), cluster--galaxy cross-correlation ($\delta_c \delta_g$), cluster auto-correlation ($\delta_c \delta_c$), and cluster lensing ($\delta_c \gamma$).}
    \label{fig:summary_of_data}
\end{figure}

In this paper, we present the cosmological constraints from galaxy clusters using the first three years of observations from the Dark Energy Survey (DES-Y3). Specifically, we jointly analyze the cluster-based data vector (\clustercomb{}\footnote{We note that this was referred to as $4\times2$pt+N in DES-Y1 \citep{4x2pt2}.} hereafter), including cluster abundances, large-scale cluster lensing, large-scale cluster--clustering, large-scale cluster--galaxy correlation functions, and large-scale galaxy--galaxy correlations, measured for DES-Y3 dataset (see Fig.~\ref{fig:summary_of_data} for a summary). As demonstrated in DES-Y1 \citep{4x2pt2}, this combination of data vectors enables efficient and robust extraction of cosmological information from galaxy clusters. Specifically, cluster–galaxy cross-correlations, cluster clustering, and galaxy clustering constrain cluster masses through the halo bias–halo mass relation, while cluster lensing provides an independent mass constraint. Together, these observables self-calibrate selection effects and yield precise cluster mass estimates. The resulting constraints on cluster mass and abundance lead to competitive cosmological constraints.

The DES-Y3 cluster sample consists of $\sim16$K  \redmapper{}\footnote{\redmapper{} stands for the red-sequence Matched-filter Probabilistic
Percolation cluster finding algorithm.} clusters across nearly $5,000$ deg$^2$, nearly tripling the sample size of DES-Y1. This increased statistical power necessitates the advancements in our modeling framework beyond DES-Y1 \citep{4x2pt1_duplicated, 4x2pt2}. Our updated analysis, validated for the precision expected in the full DES dataset \citep{Y6clustermethod}, addresses selection biases through a combination of optimized scale cuts and an improved analytic model. This approach is further validated through analytic calculations and simulations \cite{cardinal}. Additionally, we employ a specially designed machine learning-based likelihood inference tool \citep{linna}, reducing computational costs by a factor of 10. Parallel to this paper, the DES Galaxy Cluster team is working on extracting cosmology from small-scale cluster lensing while addressing systematics impacting small-scale lensing \cite{Matteoprojection, miscentering, 2021MNRAS.505...33M, Wu22, Zhang23, Salcedo23, 2023arXiv230906593A, Kelly24, Zhou24, Lee25}.

Leveraging this new cluster-based constraint alongside galaxy clustering and weak gravitational lensing measured from DES-Y3, we perform a stringent test of the \LCDM{} model. Uniquely, our cluster cosmology analysis employed a fully consistent model with DES-Y3 \ttt{}\footnote{The \ttt{} refers to the joint analyses of cosmic shear, galaxy clustering, galaxy--galaxy lensing.}, and we have homogenized analysis choices between clusters and \ttt{}. This enables relatively straightforward comparisons between cluster constraints and those from \ttt{}, as well as the joint analyses. Similar to DES-Y1, our joint analyses fully account for cross-covariance between different cosmological probes. The full data vector (\allcomb{}\footnote{We note that this was referred to as $6\times2$pt+N in DES-Y1 \citep{4x2pt2}.} hereafter) includes all data in \clustercomb{}, high redshift galaxy clustering, galaxy--galaxy lensing, and cosmic shear (See Fig.~\ref{fig:summary_of_data} for a summary). We find that adding clusters leads to $24\%$ improvements in cosmological constraints in the $\omegam-S_8$ plane.

This paper is organized as follows.
We first present the data set in Sec.~\ref{sec:data} and then the measurement and modeling in Sec.~\ref{sec:model}.
We discuss our blinding strategies in Sec.~\ref{sec:blinding}. 
The cosmological results are presented in Sec.~\ref{sec:results}. 
We conclude in Sec.~\ref{sec:conclusions}.
Appendix~\ref{sec:catalog_update} presents a DES-Y3 catalog update.
Appendix~\ref{sec:data_vec_plots} shows the full sets of data vectors.
Appendix~\ref{sec:ppd} details the calculation of the posterior predictive distribution, which has been used to quantify the goodness of fit and the tension between correlated datasets. Appendix~\ref{app:RMR} discusses the constraints on the mass--richness relation. The constraints on nuisance parameters are presented in appendix~\ref{app:all}.

\section{Dark Energy Survey data}\label{sec:data}

In this paper, we use a number of data products from the Dark Energy Survey Year 3 (DES-Y3) dataset, which comprises data taken in the first three years of DES between 2013 to 2016. The foundation of the data products described here is the DES-Y3 \texttt{Gold} catalog described in \citep{Y3gold}, from which we derive the three samples of objects: the \redmapper{} galaxy cluster sample (Section~\ref{sec:cluster_sample}), the \mcal{} source galaxy sample and the \maglim{} lens galaxy sample (Section~\ref{sec:source_lens_sample}). We note that the source and the lens galaxy samples have been described in detail in previous work \citep[see][and references therein]{Y3kp}, we therefore only summarize briefly the key aspects of the samples.
\label{sec:cluster_sample}

\subsection{DES cluster samples}
\label{sec:cluster_sample}
\begin{figure}
    \centering
    \includegraphics[width=0.5\textwidth]{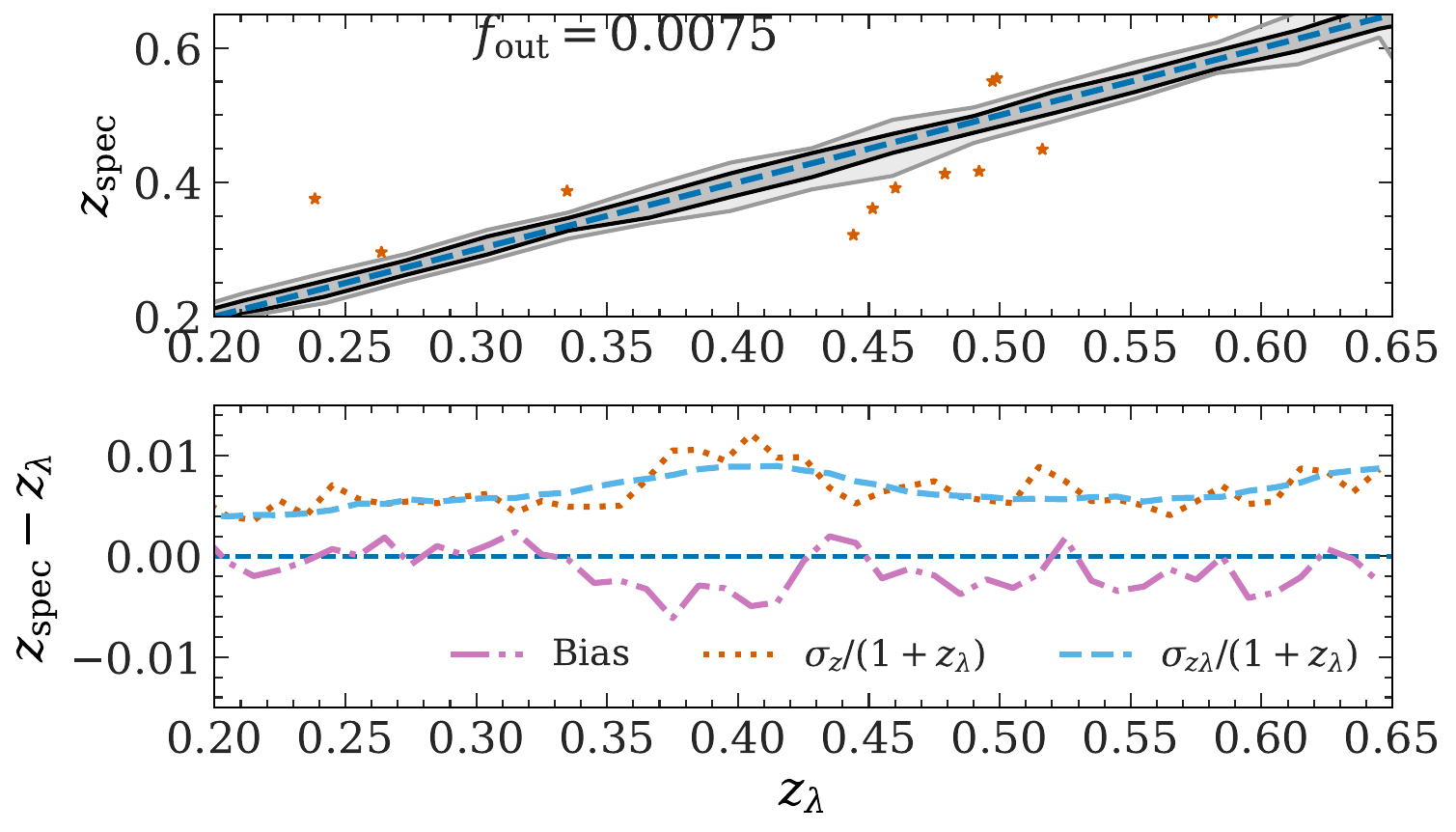}
    \caption{Photometric redshift performance of DES-Y3 \redmapper{} cluster catalog. Upper panel: gray contours show $1\sigma$ and $2\sigma$ confidence intervals, and orange stars show $5\sigma$ outliers. Bottom: photo-$z$ bias and uncertainties evaluated with available $1955$ spectroscopic central galaxies. The orange line is estimated via the standard deviation of spectroscopic redshifts. The blue line is evaluated based on the reported redshift uncertainties estimated by \redmapper{}.}
    \label{fig:redmapper_performance}
\end{figure}

For this analysis, we use a volume-limited sample of galaxy clusters detected in DES-Y3 photometric data \citep{y3-gold} with the \redmapper{} cluster finder (v6.4.22+2) \footnote{The catalog is released at \url{https://des.ncsa.illinois.edu/releases/y3a2/Y3key-cluster}.}. The \redmapper{} algorithm operates on a subset of high-quality objects selected from the DES-Y3 \texttt{Gold} catalog. To ensure data quality, we exclude objects flagged with $\rm{FLAG}\_\rm{GOLD} = 8$, $16$, $32$, or $64$. We further select extended objects using the criterion $\rm{EXTENDED}\_\rm{CLASS}\_\rm{MASH}\_\rm{SOF} \geq 2$. For photometry, we adopt the ``Single-Object Fitting'' (SOF) measurements in the $g$, $r$, $i$, and $z$ bands to identify clusters. Notably, this approach differs from DES-Y1 analyses, which relied on the multi-epoch, multi-object fitting (MOF) composite model (CM) galaxy photometry. We opt for SOF photometry in this study because it demonstrates greater stability for bright central galaxies.

\redmapper{} identifies galaxy clusters as overdensities of red-sequence galaxies. The cluster-finding process involves two main steps. First, the algorithm constructs an empirical red-sequence model, which relates galaxy colors to redshift. This model is derived using spectroscopic redshifts from the fourteenth data release of the Sloan Digital Sky Survey (SDSS DR14) \citep{SDSSDR14} and the Australian Dark Energy Survey Global Redshift Catalog (OzDES GRC). Second, the algorithm iteratively identifies overdensities of red-sequence galaxies through a matched-filter technique. The matched filter consists of galaxy colors, positions, and luminosities, which are calculated from the SOF photometry and the red-sequence model. 

Each detected galaxy overdensity, known as the \redmapper{} cluster, is then assigned a photometric redshift ($z_\lambda$), a mass proxy (richness, $\lambda$), and a central position based on the matched-filter likelihood. These properties form the basis for our subsequent cluster analyses.

\begin{figure}
    \centering
    \includegraphics[width=0.5\textwidth]{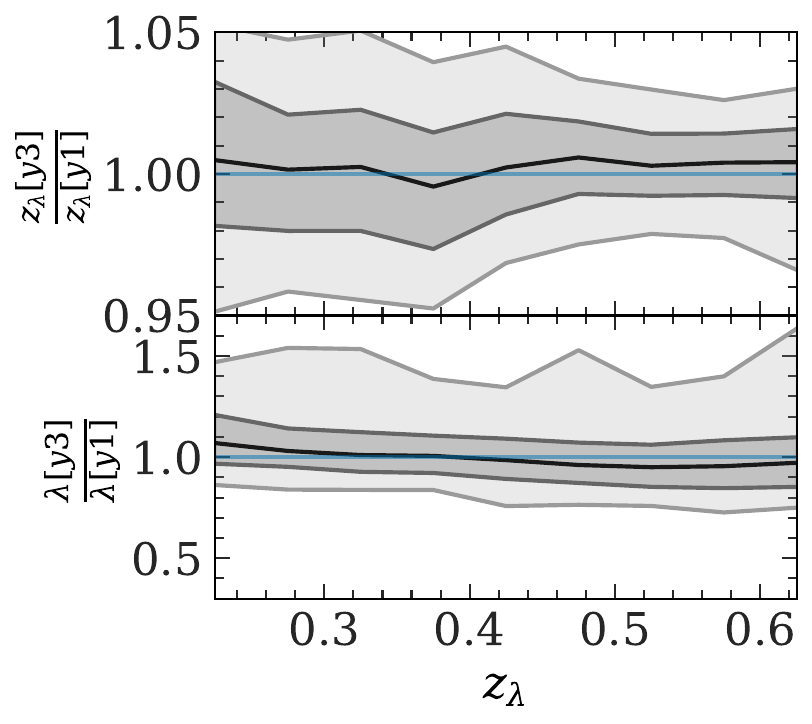}
    \caption{Comparison of DES-Y1 \redmapper{} and DES-Y3 \redmapper{} samples that have central galaxy with $<0.5$ arcmin seperation. Top panel: ratio of redshifts. Bottom panel: ratio of richness. Contours show $1\sigma$ and $2\sigma$ scatters. Black solid lines show the median.}
    \label{fig:redmappercomparison}
\end{figure}
 
In this analysis, we only use galaxy clusters with $\lambda>20$ to ensure $>99\%$ purity of the sample \citep{2016MNRAS.460.3900F, matteoSDSS}. We further restrict the samples to redshift intervals $z_\lambda=[0.2,0.65]$. The redshift lower limit mitigates the degradation of \redmapper{} performances due to the lack of $u$-band data, and the redshift upper limit ensures a relatively constant footprint across redshift and consistency of \redmapper{} redshift bins and the galaxy samples used for cross-correlations. With these restrictions, a total of 18,005 galaxy clusters are included in the DES-Y3 \redmapper{} cosmology catalog. Among $18,005$ clusters, $1,514$ of those are removed after applying a joint mask used by the \ttt{} analyses. This is $2.5$ times as many clusters compared to the cosmological sample in DES-Y1 \citep{DES_cluster_cosmology}.

The \redmapper{} v6.4.22+2 algorithm used in this analysis is similar to the one used in DES-Y1 analyses \citep{Tomclusterlensing, DES_cluster_cosmology} with one important update. The percolation radius, the radius used to deblend overlapping \redmapper{} clusters, is updated from $1.5 \times (\lambda/100)^{0.2} h^{-1}\rm{Mpc}$ to $1.95 \times (\lambda / 100) ^{0.45} h^{-1}\rm{Mpc}$ to be more consistent with the halo exclusion criteria in \cite{Tinker08}. 

We further investigate the performance of DES-Y3 \redmapper{} using available spectroscopic samples and DES-Y1 \redmapper{} cluster samples. Fig. ~\ref{fig:redmapper_performance} shows the redshift performance of the \redmapper{} samples. We compare $z_\lambda$ with the available central galaxies' spectroscopic redshift to estimate the redshift biases and scatters. In total, $1,955$ of DES-Y3 \redmapper{} clusters have a spectroscopic central galaxy, and $194$  of them have redshifts greater than $0.6$. Using these samples, we find nearly unbiased redshifts with tight scatter $\sigma(z_\lambda)/(1+z_\lambda) \simeq 0.006$, 
consistent with \citep{Tomclusterlensing}. %
Next, we compare in Fig.~\ref{fig:redmappercomparison} the richness and redshifts of \redmapper{} samples in DES-Y3 and DES-Y1 that have central galaxies within $0.5$ arcmin separation\footnote{We find a consistent result with $0.1$ arcmin.}. In this comparison, we use the DES-Y1 \redmapper{} with $\lambda \geq 5$ to avoid loss of clusters from the sample due to scattering across the $\lambda=20$ cut in DES-Y3 \redmapper{} samples. We find that the two samples have consistent redshift but slightly different richness distributions. While the median values of the redshift ratio and richness ratio between DES-Y3 and DES-Y1 are similarly consistent with $1$, we find that the scatter of the richness ratio is much more asymmetric and skewed toward larger values. This is likely due to differences in photometries and percolation radius used in the \redmapper{} algorithm.

We note that the performance of DES-Y3 \redmapper{} has been investigated in several companion papers. The centering efficiency is studied using XMM-Newton and Chandra X-ray imaging in \cite{Kelly24}. The fraction of correctly centered \redmapper{} clusters is $0.87\pm 0.04$. The distribution of radial offsets of mis-centered clusters is modeled as a Gamma distribution with a characteristic length scales of $0.23 \pm 0.05 R_\lambda$, where $R_\lambda$ is the cluster radius estimated by \redmapper{}. Refs.~\cite{2023MNRAS.522.5267U,Kelly24} investigate the X-ray temperature--richness scaling relations, finding a tight relation with a scatter of $0.23\pm0.01$ and $0.21\pm 0.01$, respectively. Further, in \citep{Grandisinprep}, the authors quantify the performance of the \redmapper{} cluster finder by cross-matching DES-Y3 clusters with detections from three South Pole Telescope surveys (SZ, pol-ECS, pol-500d). Specifically, they confirm a $\sim 20-40\%$ bias on the richness estimate due to systems in projection \citep{DES_cluster_cosmology} \footnote{We specifically test the impact of this bias in richness on cosmological constraints in \citep{Y6clustermethod}.} and rule out significant contamination by unvirialized objects at the high-richness end ($\lambda>100$).

Finally, we quantify the selection function using a customized random catalog following the method in \citep{redmappersv}. This random catalog is essential when constructing cluster-related two-point correlations. A key challenge in constructing cluster randoms is that clusters are extended objects whose detectability depends on their size, redshift, and survey boundaries. To address this, we generate \redmapper{} randoms by sampling cluster richness--redshift pairs from the data and placing them at random positions. In this process, we ensure that each cluster is assigned a location where it could have been detected based on the survey redshift mask, a footprint-dependent redshift upper limit below which all cluster member galaxies are above the detection limit. We then apply the same selection cuts as the \redmapper{} cosmology sample, removing clusters whose masked fraction exceeds $0.2$ or whose richness falls below $20$. To correct the impact of these cuts on the redshift and richness distributions, we reweight each remaining cluster with the following procedure. Each cluster richness--redshift pair is repeatedly positioned at different places within the survey footprint $1000$ times. We calculate the fraction of the $1000$ repeated samples that pass the selection criteria mentioned above. This fraction is then used as the weight for that simulated cluster. This procedure ensures that the final random catalog has a consistent selection function as the cluster cosmology sample while properly accounting for survey boundaries and depth variations.

\begin{table}
\centering
\begin{tabular}{ccc}
\multicolumn{3}{c}{\redmapper{} clusters} \\
\hline
Bin &  Redshift range & $N_\text{cluster}$ \\
\hline
1 & [0.2,  0.4] & 5,632  \\
2 & [0.4, 0.55] & 6,308  \\
3 & [0.55, 0.65] & 4,551   \\
\hline
\end{tabular}
\\
\vspace{3mm}
\begin{tabular}{cccc}
\multicolumn{4}{c}{\maglim{} galaxies} \\
\hline
Bin &  Redshift range & $N_\text{gal}$ &  $n_\text{gal}$ (arcmin$^{-2}$)  \\
\hline
1 & [0.2,  0.4] & 2,236,473 &  0.1499  \\
2 & [0.4, 0.55] & 1,599,500 &  0.1072  \\
3 & [0.55, 0.7] & 1,627,413 &  0.1091  \\
4 & [0.7, 0.85] & 2,175,184 &  0.1458  \\
\hline
\end{tabular}
\\
\vspace{3mm}
\begin{tabular}{ccccc}
\multicolumn{4}{c}{\mcal{} source galaxies} \\
\hline
Bin  & $N_\text{gal}$ &  $n_\text{eff}$ (arcmin$^{-2}$) & $\sigma_\epsilon$ \\
\hline
1 & 24,940,465 &  1.476 & 0.243 \\
2 & 25,280,405 &  1.479 & 0.262 \\
3 & 24,891,859 &  1.484 & 0.259 \\
4 & 25,091,297 &  1.461 & 0.301 \\
\hline
\end{tabular}
\caption{Basic characteristics of the source galaxy samples, lens galaxy samples, and cluster samples. The cluster sample has three tomographic bins, while each galaxy sample has four tomographic bins. For the lenses, we list the redshift range, total galaxy number counts, and number density. For the sources, we list the total number of galaxy counts, as well as the effective number density and shape noise for weak lensing. The area of the survey is 4,143 deg$^2$. }
\label{tab:gal_samples}
\end{table}

\subsection{DES source and lens galaxy samples}
\label{sec:source_lens_sample}

We use the same source and lens galaxy samples as those used in \citep{Y3kp} (see Appendix~\ref{sec:catalog_update} for a minor update to the source catalog). Using the same sample is the key to cleanly and coherently combining the cluster information with the \ttt{} information. 

The \mcal{} source sample is derived from the \mcal{} algorithm \citep{Sheldon2017} and rigorously examined in \citep{Gatti+2022}. The final catalog consists of $\sim100$M galaxies divided into 4 tomographic bins. The weighted source number density is $n_{\rm eff}=5.59$ gal/arcmin$^2$, with a corresponding shape noise of $\sigma_e=0.261$. The redshift distribution and its calibration using independent methods based on photometry as well as clustering information is described in \citep{y3-sompz, y3-sourcewz}. Using image simulations, \citep{y3-imagesims} quantified the uncertainty in the shear calibration as well as its coupling with the redshift distribution due to blending.

The \maglim{} lens sample is constructed via a redshift-dependent magnitude selection from the DES-Y3 \texttt{Gold} catalog and is designed to have the maximum statistical power while maintaining control over the redshift uncertainties \citep{y3-2x2maglimforecast}. To minimize spurious clustering coming from spatially varying systematic effects, \citep{y3-galaxyclustering} derive a large-scale structure weight that is included with the catalog. The definition of the bins, as well as the redshift distribution and its uncertainty, are derived using the Directional Neighborhood Fitting (DNF) algorithm \citep{DeVicente2016}. The original sample includes $6$ tomographic bins. In \citep{Y3kp}, only four out of six redshift bins were used in the final cosmology analysis due to poor fits in the high-redshift bins. In this work, we further exclude the highest-redshift bin (bin 4) when cross-correlating with the cluster sample due to the lack of overlap in redshift. 

\begin{figure}
    \centering
    \includegraphics[width=0.4\textwidth]{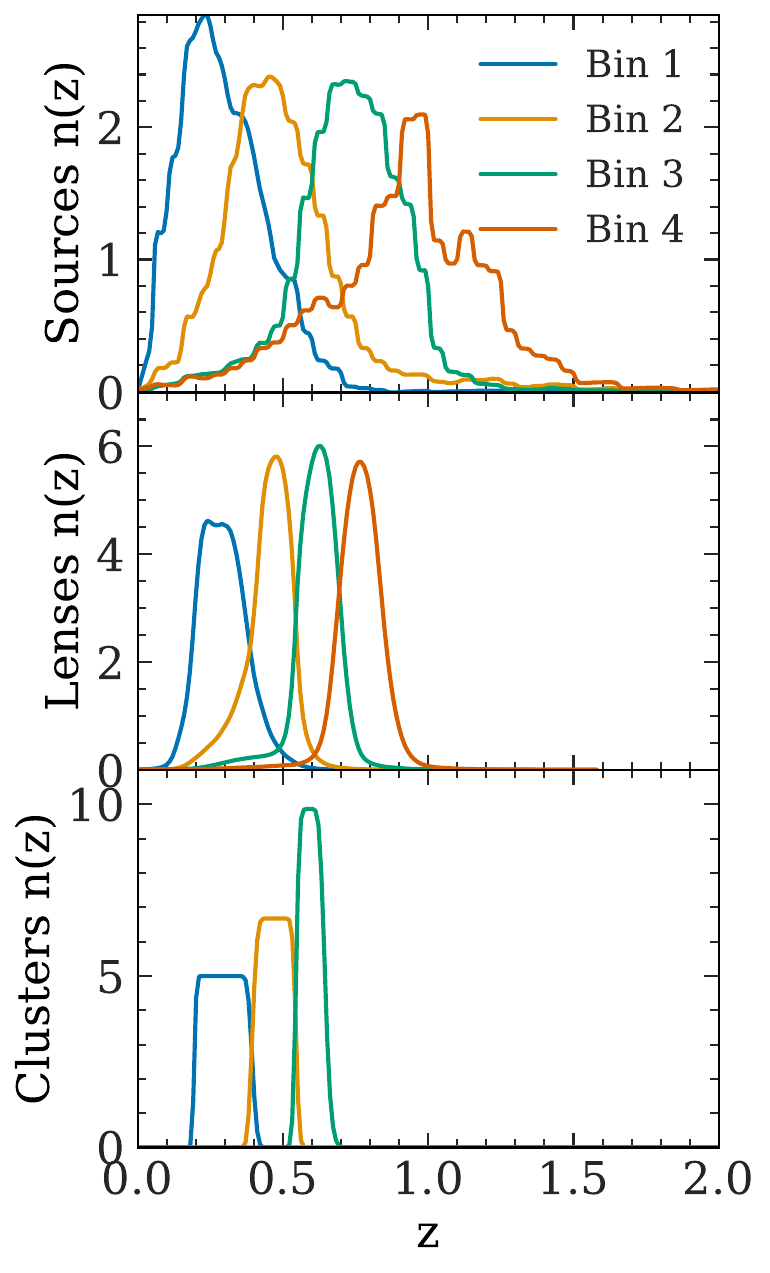}
    \caption{The \mcal{} source galaxy (top), \maglim{} lens galaxy (middle), and \redmapper{} cluster (bottom) redshift distributions. }
    \label{fig:redshift}
\end{figure}

Table~\ref{tab:gal_samples} lists the key characteristics of the source and lens sample, while Fig.~\ref{fig:redshift} shows the redshift distribution of the samples. Constraints on the shear and redshift calibration parameters are listed in Table~\ref{tab:params}. 

\begin{table*}
    \centering
    \footnotesize
    \caption{Parameters and priors considered in this analysis. ``Flat'' represents a flat prior in the given range, ``Gauss $(\mu, \sigma)$'' denotes a Gaussian prior with mean $\mu$ and width $\sigma$, and ``Fixed $(X)$'' means that the parameter is fixed at $X$.\\}
    \label{tab:params}
   \begin{tabular}{ lccc}
    \hline 
	Parameter	 & Prior & Varied in \clustercomb{} & Varied in \allcomb{} \\
	\hline 
	\textbf{Cosmology} && \\
	$\omegam$ & Flat (0.1,0.9) & \checkmark{} &\checkmark{}\\
	$A_s \times 10^9$ & Flat (0.5, 5.0) &  \checkmark{} &\checkmark{}\\
	$n_s$ & Flat (0.87, 1.07) &  \checkmark{} &\checkmark{}\\
	$\Omega_{\rm b}$ &  Flat (0.03, 0.07) & \checkmark{} &\checkmark{}\\
	$h$ &  Flat (0.55, 0.91)& \checkmark{} &\checkmark{} \\
    $\Omega_{\nu}h^2$ &  Flat (0.0006, 0.00644) &  \checkmark{} &\checkmark{}\\
	\hline 
	\textbf{Galaxy Bias} && \\
	$b_{1,\rm{l}}^1$ &  Flat (0.8, 3.0) & \checkmark{} &\checkmark{}\\ 
	$b_{1,\rm{l}}^2$ &  Flat (0.8, 3.0) & \checkmark{} &\checkmark{}\\ 
	$b_{1,\rm{l}}^3$ &  Flat (0.8, 3.0) & \checkmark{} &\checkmark{}\\ 
    $b_{1,\rm{l}}^4$ &  Flat (0.8, 3.0) & - &\checkmark{}\\ 
	\hline
 \textbf{Intrinsic Alignment}&& \\
 $a_1$ & Flat (-5.0, 5.0)&  \checkmark{} &\checkmark{} \\
  $\eta_1$ & Gauss $(0, 3)$  &  \checkmark{} &\checkmark{} \\
  $a_2$ & Flat (-5.0, 5.0)&  - &\checkmark{} \\
  $\eta_2$ & Gauss $(0, 3)$ & -&\checkmark{}  \\
  $b_{\rm TA}$ & Fixed (1) & - &-\\
 \hline
	\textbf{\maglim{} Photo-$z$}&&\\
	$\Delta_{z,\rm{l}}^1 \times 10^2$ & Gauss $(-0.9, 0.7)$& \checkmark{} &\checkmark{} \\
	$\Delta_{z,\rm{l}}^2\times 10^2$ & Gauss $(-3.5, 1.1)$& \checkmark{} &\checkmark{}\\
	$\Delta_{z,\rm{l}}^3\times 10^2$ & Gauss $(-0.5, 0.6)$& \checkmark{} &\checkmark{}\\
	$\Delta_{z,\rm{l}}^4\times 10^2$ & Gauss $(-0.7, 0.6)$& -&\checkmark{}\\
    $w_{z,{\rm l}}^{1}$ & Gauss (0.98, 0.06) &\checkmark{}&\checkmark{} \\
    $w_{z,{\rm l}}^{2}$ & Gauss (1.31, 0.09) & \checkmark{}&\checkmark{} \\
    $w_{z,{\rm l}}^{3}$ & Gauss (0.87, 0.05) & \checkmark{}&\checkmark{} \\
    $w_{z,{\rm l}}^{4}$ & Gauss (0.92, 0.05) & - & \checkmark{}\\
	\hline
    \textbf{\maglim{} Magnification}&&\\
	$C_{\rm{l}}^1$ & Fixed (0.42)&  - &-\\
	$C_{\rm{l}}^2$ & Fixed (0.30)&  - &-\\
	$C_{\rm{l}}^3$ & Fixed (1.76)& - &- \\
        $C_{\rm{l}}^4$ & Fixed (1.94)&  - &-\\
    	\hline
\textbf{Point-mass Marginalization}&&\\
	$B^i$ & Flat (-1.0, 1.0) &  - &\checkmark{}\\
	\hline
	\textbf{Source Galaxy Photo-$z$} && \\
	$\Delta_{z,\rm{s}}^1 \times 10^2$& Gauss $(0.0, 1.8)$ & \checkmark{} &\checkmark{}\\
	$\Delta_{z,\rm{s}}^2 \times 10^2$ & Gauss $(0.0, 1.5)$ & \checkmark{} &\checkmark{}\\
	$\Delta_{z,\rm{s}}^3 \times 10^2$& Gauss $(0.0, 1.1)$ &  \checkmark{} &\checkmark{}\\
	$\Delta_{z,\rm{s}}^4 \times 10^2$& Gauss $(0.0, 1.7)$ &  \checkmark{} &\checkmark{}\\
		\hline
\textbf{Shear Calibration}& &\\
         $m_1 \times 10^2$ & Gauss $(-0.6, 0.9)$ & \checkmark{} &\checkmark{} \\
$m_2 \times 10^2$ &  Gauss $(-2.0, 0.8)$ & \checkmark{} &\checkmark{}\\
$m_3 \times 10^2$ & Gauss $(-2.4, 0.8)$& \checkmark{} &\checkmark{} \\
$m_4 \times 10^2$ & Gauss $(-3.7, 0.8)$ & \checkmark{} &\checkmark{}\\
 \hline
    \textbf{\redmapper{} Richness--Mass Relation} && \\
    $\rm{ln}\lambda_{0}$ &  Flat (2.0,5.0) & \checkmark{} &\checkmark{}\\
    $A_{\rm{ln}\lambda}$ &  Flat (0.1,1.5) & \checkmark{} &\checkmark{}\\
    $B_{\rm{ln}\lambda}$ & Flat (-5.0, 5.0) & \checkmark{} &\checkmark{}\\
    $\sigma_{\rm{intrinsic}}$ &  Flat (0.1, 1.0) & \checkmark{} &\checkmark{}\\
    \hline 
    \textbf{\redmapper{} Selection Effect} && \\
    $b_{s1}$ &  Flat (1.0,2.0) & \checkmark{} &\checkmark{}\\
    $b_{s2}$ & Flat (-1.0,1.0) & \checkmark{} &\checkmark{}\\
    $r_0$ & Flat (10, 60) &  \checkmark{} &\checkmark{}\\
    \hline
    \textbf{\redmapper{} Magnification}&&\\
	$C_{\rm{c}_A}^i$ & Fixed (-2)& -&-\\
	\hline
    \end{tabular}
\end{table*}

\section{Data vectors and modeling strategy}
\label{sec:model}
\subsection{Measurements}

We split the sample of 16,491 DES-Y3 clusters within the DES-Y3 3$\times$2pt footprint into three redshift bins in the range $0.2 < z < 0.65$. Within each tomographic bin, the clusters are further split into four richness bins, $20\le \lambda <30,\, 30\le \lambda <45,\, 45\le \lambda < 60,\, 60 \le \lambda $. The corresponding number counts $N$ are shown in Appendix \ref{sec:data_vec_plots}; these 12 data points have a combined signal-to-noise (S/N) of $94.5$. For all summary statistic measurements, presented in the figures shown in Appendix \ref{sec:data_vec_plots}, each panel shows the data points in the upper part and the fractional difference between the data and the mean of the predictions from the \clustercomb{} chains, normalized by the PPD prediction scatter (Appendix~\ref{sec:ppd}). Data points excluded from the analysis by scale cuts are shown in light opacity.

We use the \textsc{TreeCorr} code \citep{treecorr} to measure two-point auto- and cross-correlation functions of the DES-Y3 cluster sample, the \maglim{} lens sample, and the \mcal{} source sample. As mentioned in Sec.~\ref{sec:source_lens_sample}, the \clustercomb{} analysis restricts the lens galaxy sample to \maglim{} tomography bins 1--3, as \maglim{} bin 4 does not overlap with the cluster redshift range  $0.2 < z < 0.65$, and we do not detect a cross-correlation at the sensitivity in this analysis.

The resulting two-point statistics are $\gamma_{\rm{c}}(\theta)$, the tangential shear profile per cluster richness bin and cluster-source tomography bin combination (with 404 data points after scale cuts and a S/N of $31.8$), $w_{\rm{cc}}$, the angular clustering of clusters across richness bins within each tomography bin (with 149 data points after scale cuts and a S/N of $18.8$), and $w_{\rm{cg}}$, the angular cross-clustering of clusters and galaxies per cluster richness bin and cluster-galaxy tomography bin combination (with 124 data points after scale cuts and a S/N of $39.6$). We use measurements of the \maglim{} angular correlation function $w_{\rm{gg}}$ from \citep{y3-galaxyclustering} (31 data points after scale cuts and a S/N of $52.5$), which we reproduce in the figure presented Appendix~\ref{sec:data_vec_plots} to illustrate the fractional difference between $w_{\rm{gg}}$ and the mean of the predictions from the \clustercomb{} chains. All two-point measurements are presented in Appendix \ref{sec:data_vec_plots}.

\subsection{Modeling Strategy}
The theoretical model for the \clustercomb{} and \allcomb{} analyses is described in detail and validated in \citep{Y6clustermethod}, building on the 3$\times$2pt model \citep{y3-generalmethods, Y6Model}. Briefly, the 3$\times$2pt model is based on a model for the non-linear matter power spectrum, linear galaxy bias, the tidal alignment tidal torquing (TATT) intrinsic alignment model \citep{TATT}, which is an extension of the non-linear alignment (NLA) model. The massive neutrinos are modeled as three degenerate species of equal mass. Our model further included magnification of lens galaxies, photometric redshift uncertainties parameterized by a shift parameter (\mcal{}) or shift and stretch parameters (\maglim{}), and multiplicative shear calibration uncertainty; non-local contributions to galaxy--galaxy lensing from the mass distribution below the scale cut are marginalized out, which we implement as a parametric marginalization (``point mass''). 
The theoretical predictions for the cluster observables are calculated using a log-normal richness--mass relation and an empirical, scale-dependent model for \redmapper{} selection effects (see section C5 of \cite{Y6clustermethod} for details). The model for cluster 2pt-statistics ($\gamma_{\rm c}$, $w_{\rm cc}$, $w_{\rm cg}$) is an extension of the 3$\times$2pt model, with the linear bias of each cluster richness bin computed from the observable--mass relation and the halo bias--mass relation.  The cluster lensing goes through a linear transformation based on the relation of the tangential shear profile ($\Delta\Sigma$)  and the projected surface density ($\Sigma$) to localize the signal \citep{Ytransform}.  Scale cuts for cluster (cross-) clustering and cluster lensing are determined to control the impact of non-linear biasing and uncertainties in the modeling of the non-linear matter distribution.

The model parameters and priors are summarized in Table~\ref{tab:params}. We note that compared to the DES-Y3 3$\times$2pt analysis presented in \citep{Y3kp}, we implement two changes for consistency with the upcoming DES Year 6 analyses: the matter power spectrum model is updated to \textsc{HMCode2020} \citep{HMcode2020}, and we employ weakly informative priors on the redshift evolution of intrinsic alignments ($\eta_1,\eta_2$) to reduce prior volume effects.

Simulated analyses of noise-less model vectors indicate that marginalized parameter constraints should only be weakly affected by prior volume effects, with the 2D marginalized constraint on $\omegam$ and $S_8$ biased by less than $0.3$ of the statistical uncertainties.

The likelihood inferences are performed using a customized sampler LINNA \citep{linna}. LINNA automatically builds a theory emulator, iteratively modifies the training sample, and performs MCMC analyses. The accuracy of LINNA for \ttt{}, \clustercomb{}, and \allcomb{} has been validated to the expected constraining power of LSST-Y10. 

\section{Blinding and unblinding}
\label{sec:blinding}

In this paper, the \clustercomb{} portion of the analysis is done in a blinded fashion to avoid any implicit decisions based on the results from the data. The blinding and unblinding protocol was defined before making any measurements with data and followed to minimize any unintentional analysis decisions being affected by the data results. The philosophy follows what was done in \citep{Y3kp} and \citep{To2021}. Below, we describe the blinding strategies, the findings during unblinding, and any changes in the analysis after unblinding. 

\subsection{Blinding}

As this analysis is done after the 3$\times$2pt analysis is unblinded \citep{Y3kp}, there is no catalog or data vector-level blinding. We only perform blinding at the parameter level. That is, we run chains directly on the unblinded data vectors, but the output chain samples are shifted before being saved and analyzed.
 
For the cosmological parameters of interest ($\omegam$, $h$, $\Omega_{\rm b}$, $n_s$, $A_s$, $\sum m_{\nu}$), we apply a random shift drawn from a uniform distribution with an upper limit of 5$\sigma$ of the posterior of that parameter and a lower limit of 0. For the 4 mass--observable relation parameters ($\ln \lambda_0$, $A_{\ln \lambda}$, $B_{\ln \lambda}$, $\sigma_{\rm intrinsic}$) and the 3 galaxy bias parameters (only the first three lens bins were used in the \clustercomb{} part), we apply the same procedure but with an upper limit of 2$\sigma$ of the posterior. 

Note that we do not blind all other parameters or the $\chi^2$ values. We are allowed to plot the unblinded data vector and best-fit model without blinding, as well as the blinded contours. 

\subsection{Unblinding}
\label{section:unblind}
To unblind, we have defined a list of tests that need to be passed. There are three main categories of tests that we describe below:
\begin{itemize}
\item \textbf{Modeling tests:} These tests check that we can recover unbiased cosmology with our modeling choices. In particular, they verify that with the scales used in the analysis, we can recover unbiased cosmology even with uncertainty in some of the modeling choices. Most of these tests are carried out and thoroughly checked in \citep{Y6clustermethod} using simulated data vectors. In this paper, we conduct one additional test of the model. 

\begin{itemize}

\item \textit{Redshift-dependent selection effect.} We investigate whether the redshift evolution of the selection effect needs to be explicitly modeled. To do this, we introduce a redshift-dependent parameterization of the selection effect amplitude, $b_{s1}$ (for the equation of the selection effect model, see equation 23 of \citep{Y6clustermethod}):

\begin{equation}
    b_{s1} \left(\frac{1+z}{1.45}\right)^{b_z},
\end{equation}
where $b_z$ is a free parameter governing the redshift evolution. If $b_z$ is consistent with $0$ within $3\sigma$, we do not consider a redshift-dependent selection effect model.	Repeating our \clustercomb{} cosmological analysis with this modification, we obtain a marginalized $1\sigma$ constraint of $b_z = -0.04^{+0.29}_{-0.34}$, consistent with zero.  The resulting cosmological constraints remain consistent with our fiducial analysis, as shown in the second row of Fig.~\ref{fig:4x2pt_table}. This result indicates that redshift evolution in the selection effects is negligible and has minimal impacts on cosmological constraints. \\
\end{itemize}

\item \textbf{Data-level tests:} These tests empirically examine whether there are any unexpected data behaviors. All the tests are thus run directly on the data itself. There are a number of tests here that we summarize. Overall, our results are summarized in Fig.~\ref{fig:4x2pt_table}, and we find that there are no significant issues in the data that prevent us from unblinding. 
\begin{itemize}

\item \textit{Systematics weights.} We test the effect of varying survey conditions, which might imprint an artificial clustering signal on the large-scale two-point correlation functions. We expect this effect to be negligible for the \clustercomb{} analysis due to the reasons below. First, cross-correlations such as cluster--galaxy cross-correlations and cluster lensing have much higher signal-to-noise than clustering of clusters. As long as the survey systematics on the galaxy sample are removed, the cross-correlations are immune from this systematics even if we do not correct the impact on the cluster density field. Second, the cluster randoms are constructed by injecting fake clusters on real data and rerunning the detection. This process is expected to remove most of the survey systematics for the relative density of clusters and randoms. To validate these expectations, we conduct an explicit test using DES-Y3 data. We first match \redmapper{}, \redmapper{} randoms, and broad-$\chi^2$ \redmagic{} samples \citep{y3-2x2ptbiasmodelling} in DES-Y3 by their positions. The broad-$\chi^2$ \redmagic{} sample, one of the two lens galaxy samples in the DES-Y3 \ttt{} cosmology analyses, is selected for its color consistency with the \redmapper{} red sequence model. We apply its systematic weights, which correct survey systematics in the \redmagic{} galaxy density fields, to both the \redmapper{} sample and its randoms. Using these weights, we generate a new data vector and compare the resulting cosmological constraints to those from our fiducial analysis. Because galaxies should be more affected by survey systematics than galaxy clusters, the difference between the two analyses sets an upper bound on the impact of systematics on the clustering signal for \clustercomb{}. Since we expect negligible changes in cosmological constraints, our requirement for this test is that $S_8$ and $\omegam$ constraints should shift within $0.3\sigma$. We find that the difference in $S_8$ and $\omegam$ constraints between the two analyses is $0.068\sigma$, confirming that survey systematics have a negligible effect.\\

\item \textit{Cluster lensing estimator.} 
Our analysis choices for cluster lensing are different from those of DES-Y1. We apply a linear transform \citep{Ytransform} to localize the cluster lensing signal \citep{Y6clustermethod}, and we adopt a scale cut ($2~h^{-1}\rm{Mpc}$ on the transformed cluster lensing) that is different from Y1 ($8~ h^{-1}\rm{Mpc}$ on $\gamma_c$). We first test our analysis with DES-Y1 analysis choices \citep{4x2pt1_duplicated, 4x2pt2}, where we do not perform localized transform but adopt $8~\hMpc$ as our scale cut. We then assess whether including small-scale cluster lensing biases the cosmological constraints by repeating the analysis with the cluster lensing signal removed at $2$–$5~h^{-1}\rm{Mpc}$. Since it is difficult to define the requirements for these tests, we qualitatively examine the posterior. If the shift in the $\omegam$-$S_8$ plane is greater than $3 \sigma$ of the fiducial analysis, we investigate further. 

As shown in Fig.~\ref{fig:4x2pt_table}, we find that Y1 analysis choices lead to higher $\omegam$ (mean $\omegam=0.27$ to $0.28$) but lower $\sigma_8$ (mean $\sigma_8=0.92$ to $0.87$). Ref.~\citep{Ytransform} have shown that galaxy--galaxy lensing measurements contain one-halo contribution even at $12~\hMpc$. Because structures near clusters are more non-linear, we expect that cluster lensing contains even more one-halo contribution at $12~\hMpc$. The Y1 analysis likely overestimates the lensing signal due to this residual one-halo contribution, leading to a bias toward higher $\omegam$ and correspondingly lower $\sigma_8$ to maintain the same cluster abundance. While adopting the DES-Y1 analysis choices does not fully shift the DES-Y3 cosmological constraints to match DES-Y1 results \citep{4x2pt2}, the trend is consistent. The remaining difference is well within expectations from statistical fluctuations. Additionally, removing small-scale cluster lensing only mildly shifts the cosmological constraints, highlighting the robustness of our constraints to small-scale systematics once the cluster lensing signal is localized.\\
\begin{figure*}
    \centering
    \includegraphics[width=0.85\textwidth]{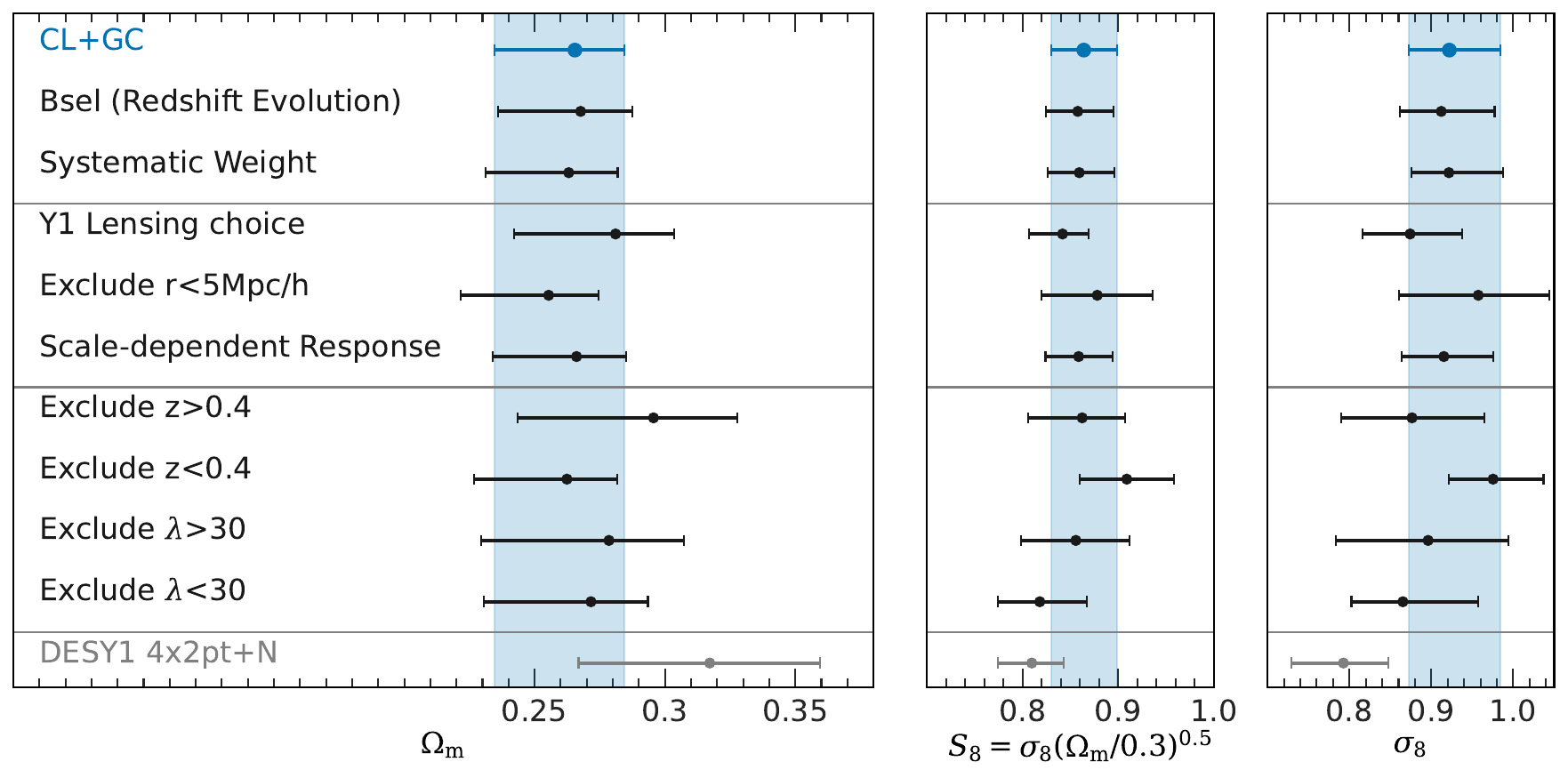}
    \caption{Summary of marginalized constraints (mean and $1\sigma$ confidence interval) on $S_{8}$, $\omegam$, and $\sigmae$ in \LCDM. The first section shows additional modeling tests beyond those presented in \citep{Y6clustermethod}. The second section shows the impact of cluster lensing analysis choice, while the third section shows the consistency of different data splits. The final section shows the constraints from DES-Y1 data as a comparison.}
    \label{fig:4x2pt_table}
\end{figure*}

\item \textit{Scale-dependent} \mcal{} \textit{response.} 
We check whether using scale-dependent \mcal{} responses impacts our results. While the analyses of the South Pole Telescope (SPT) \citep{SPTcosmo} and eROSITA \citep{erassmass} adopt a scale-dependent response, we choose to adopt a scale-independent response as our fiducial analysis choice. This is because it is unclear whether the measured scale-dependent response is due to the contribution of cluster member galaxies, which do not contribute to the lensing signal. Adopting a scale-dependent response could lead to a bias in cluster lensing measurement. Since it is unclear whether a scale-dependent response should be adopted, it is difficult to define the requirements for this test. Our threshold for unblinding is that the shift in the $\omegam$-$S_8$ plane between analyses with and without scale-dependent responses is less than $3 \sigma$ of the fiducial analysis. We show in Fig.~\ref{fig:4x2pt_table} that this analysis choice leads to negligible impacts on our cosmological constraints. \\
 
\item \textit{Data split test:} we perform the cosmological inference with a subset of the data to check for consistency.  In particular, we split the cluster sample according to redshift ($z<0.4$ and $z>0.4$) and richness ($\lambda<30$ and $\lambda>30$). We also split the two-point data vector into subsets that contain lensing and clustering. Since it is difficult to define the requirements for this test, we qualitatively examine the posterior. Our threshold for unblinding is that the shift in the $\omegam$-$S_8$ plane between analyses of two subsets of the data is less than $3 \sigma$ of the fiducial analysis. As shown in Fig.~\ref{fig:4x2pt_table}, all splits yield consistent cosmological constraints, reinforcing the robustness of our result.
\item \textit{Covariance matrix:} we check that the implementation of the shape noise component in the covariance is consistent between an analytical calculation and that from randomly rotating the galaxies. \\

\end{itemize}

\item \textbf{Goodness-of-fit tests:} 
We want to test whether our model is a good description of the data. We have to predefine a course of action in the scenario that our model does not fit the data so that we do not make decisions in favor of the model we considered. In particular, we use the posterior predictive distribution 
(PPD) methodology described in Appendix~\ref{sec:ppd} to evaluate the goodness-of-fit in a fully Bayesian way. We set the threshold for unblinding to be 0.01. 

\end{itemize}

Once all the tests above were passed, we unblind the cosmological constraints for \clustercomb{}. There have been no changes in the analysis after unblinding.

\section{Cosmological constraints}
\label{sec:results}

We employ DES cluster measurements and a thorough analysis pipeline presented in Sec.~\ref{sec:model} to test the currently favored \LCDM model (the 5-parameter $\Lambda$CDM model with a varying neutrino mass). One unique aspect of DES-Y3  cluster constraints is that we homogenize analysis choices and systematic models between DES-Y3 clusters and DES-Y3 \ttt{} cosmological analyses, enabling an apples-to-apples comparison between different cosmological probes and, eventually, joint analyses. In this section, we first discuss the robustness of DES cluster cosmology results and compare them with other optical, X-ray, and millimeter cluster cosmology constraints. We then discuss the consistency between DES clusters and other DES cosmological probes. Finally, we present results from joint analyses of clusters and \ttt{} and compare them with cosmology from external datasets. Throughout this section, we use the PPD methodology to assess goodness-of-fit and evaluate consistency between different data vectors within a fully Bayesian framework (see Appendix~\ref{sec:ppd} for details). A low PPD value (e.g., $0.01$ \footnote{Note that this is the unblinding criteria of the DES-Y3 \ttt{} analysis  \citep{Y3kp}.} ) signals potential inconsistencies between the model and data or among different datasets.

\subsection{DES-Y3 \clustercomb{} cluster cosmology}

\begin{figure}
    \centering
    \includegraphics[width=0.35\textwidth]{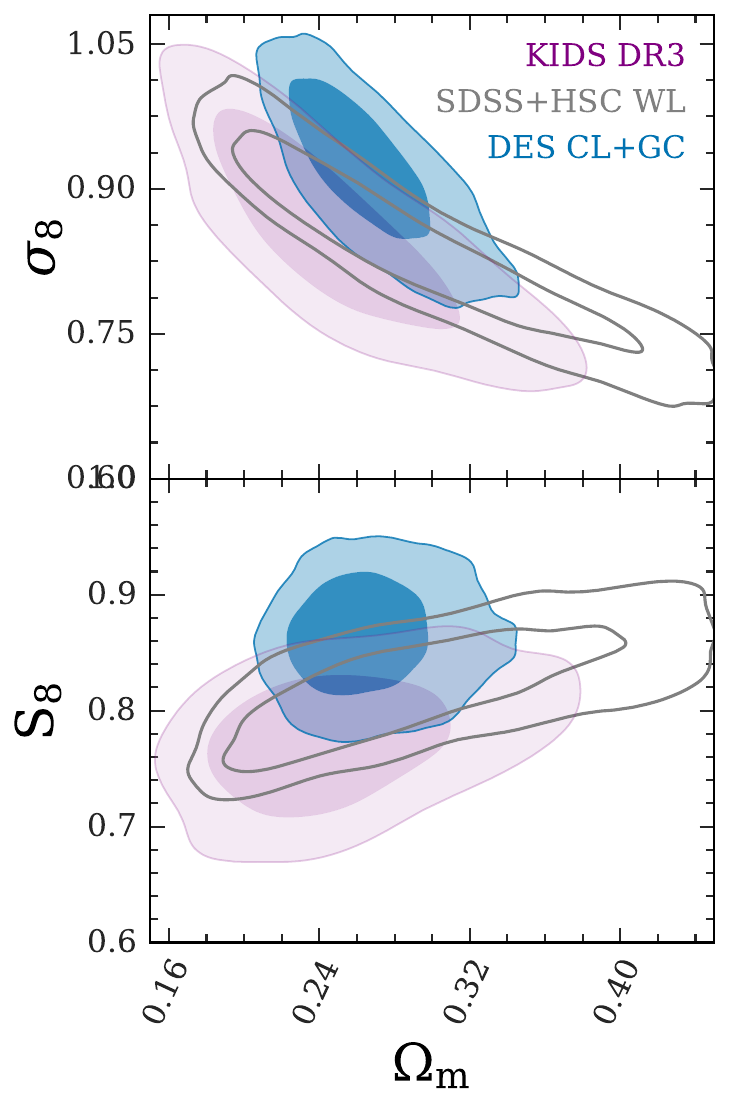}
    \caption{Marginalized constraints on the three parameters $\sigma_8$, $S_8=\sigma_8\sqrt{\omegam/0.3}$, and $\omegam$ in the \nuCDM{} model from Stage-III optical cluster cosmology analyses. Contours show $68\%$ and $95\%$ confidence intervals.}
    \label{fig:opticalcluster}
\end{figure}

\begin{table*}
\centering
\begin{tabular} { l  c c c c}
\noalign{\vskip 3pt}\hline\noalign{\vskip 1.5pt}\hline\noalign{\vskip 5pt}
 
 Parameter & \ttt{}  & \clustercomb{}  & \allcomb{}  & \allcomb{}+Planck CMB\\
\hline
{\boldmath$\omegam         $} & $0.332^{+0.032}_{-0.042}   $ & $0.265^{+0.019}_{-0.031}   $ & $0.294^{+0.022}_{-0.033}   $ & $0.317^{+0.007}_{-0.011} $\\

{\boldmath$A_s (\times 10^{-9})             $} & $1.988^{+0.232}_{-0.442}$ & $2.527^{+0.323}_{-0.544}$ & $2.068^{+0.249}_{-0.450}$ & $2.092^{+0.028}_{-0.033}$\\

{\boldmath$\sum m_\nu$ (eV)} & -                  & -            & - & $< 0.258\ (95\% \rm{CL})                  $\\

{\boldmath$h               $} & -& - & - & $0.672^{+0.008}_{-0.006}$\\

{\boldmath$\sigma_8        $} & $0.748^{+0.053}_{-0.063}   $ & $0.922^{+0.063}_{-0.049}   $ & $0.822\pm 0.053            $ & $0.790^{+0.016}_{-0.010}   $\\

{\boldmath$S_8             $} & $0.784\pm 0.022            $ & $0.864\pm 0.035            $ & $0.811^{+0.022}_{-0.020}   $ & $0.812^{+0.012}_{-0.011}   $\\
\hline
\end{tabular}
\caption{Summary of the marginalized parameter constraints in $\nuCDM$. The mean and $68\%$ confidence interval are provided for each cosmological parameter. Parameters that are not constrained are indicated by a dash.}
\label{tab:paramsum}
\end{table*}

In Fig.~\ref{fig:opticalcluster}, we show the marginalized \clustercomb{} constraints from the DES-Y3 \redmapper{} clusters for $\sigma_8$, $S_8$, and $\omegam$. The numerical values of the constraints are shown in Table~\ref{tab:paramsum}. We find that $\nuCDM$ well-describes DES-Y3 cluster measurements. Using the PPD metric \citep{y3-inttensions} to quantify the goodness-of-fit (see Appendix~\ref{sec:ppd}), we find $p(\rm{CL}+\rm{GC} | \nuCDM)=0.39$. %
Marginalized over $28$ astrophysical parameters, \clustercomb{} constraints on the key parameters are 
\begin{equation}
  \begin{aligned}
    S_8              &= 0.864 \pm 0.035 \\
    \omegam          &= 0.265^{+0.019}_{-0.031} \\
    \sigma_8         &= 0.922^{+0.063}_{-0.049} .
  \end{aligned}
\end{equation}
The figure-of-merit\footnote{The figure-of-metrit is calculated as $1/\sqrt{\rm{det}(\rm{Cov}(S_8,\omegam))}$.} on $\omegam$--$S_8$ of DES-Y3 \clustercomb{} to DES-Y1 \clustercomb{} is $1.52$, which is expected by the improved statistical power \citep{4x2pt2}.

We now compare our results with other optical cluster cosmology analyses in Fig.~\ref{fig:opticalcluster}. Ref.~\citep{Sunayama2023} calibrate \redmapper{} clusters detected in SDSS using HSC-Y3 weak lensing data, obtaining constraints that are broader but consistent with our \clustercomb{} results. Similarly, Ref.~\citep{KIDScluster} analyzes optically selected clusters in the KiDS survey using KiDS-DR3 weak lensing. While their constraints are also broader and consistent with DES \clustercomb{}, they find a slightly lower value of $S_8$. While we focus on comparisons with the latest results from optical clusters, we show comparisons of various DES cluster cosmology analyses in Appendix~\ref{app:descompare}.

Fig.~\ref{fig:allcluster} extends this comparison to cluster cosmology constraints from different wavelengths. Ref.~\citep{SPTcosmo} analyze clusters detected in the SPT-SZ and SPTpol surveys, with mass calibration performed using DES-Y3 and HST weak lensing. Ref.~\citep{erasscluster} study clusters from the western Galactic hemisphere of eROSITA's first All-Sky Survey (eRASS1), calibrating masses with DES-Y3, KiDS, and HSC weak lensing datasets \citep{erassmass, erasskids}. The figure-of-merit on $\omegam$--$S_8$ of DES-Y3 \clustercomb{} to SPT and eRASS is $0.75$ and $0.2$ respectively. We find that our constraints are consistent with eRASS1 and SPT. The central value of $S_8$ from DES-Y3 clusters is similar to that of eRASS1 and is somewhat higher than SPT. Interestingly, the mean mass of DES-Y3 clusters is more similar to that of eRASS1 than SPT (see Fig.~\ref{fig:MOR}), and the redshift range of eRASS1 ($z=0.1-0.8$) and DES-Y3 ($z=0.2-0.65$) is more similar than SPT ($z=0.25-1.78$). 
The consistent deviation of DES-Y3 and eRASS1 from SPT could suggest a mass-dependent or redshift-dependent trend in $S_8$ constraints derived from galaxy clusters. However, given the current statistical precision of DES-Y3, this remains an intriguing possibility rather than a definitive conclusion.

The DES-Y3 cluster cosmology analysis differs in key ways from most other cluster studies. For example, KiDS, HSC, eRASS, and SPT all rely on cluster lensing at scales below $2 ~\hMpc$ for mass calibrations, while DES-Y3 clusters remove those scales for analyses. DES-Y3 clusters uniquely incorporate cluster--galaxy cross-correlations for mass calibrations. DES-Y3 clusters consider the full modeling complexities of DES weak lensing analysis, while others simplify some of the modeling choices, such as intrinsic alignment, magnifications, etc., although we note that our analysis approach is slightly more sensitive to these effects. Despite these differences, the level of consistency between DES-Y3 cluster cosmology and results from independent optical, X-ray, and SZ-selected cluster analyses is remarkable.
This agreement, across diverse data sets and modeling assumptions, highlights the reliability and the great potential of galaxy clusters as a cosmological probe.

\begin{figure}
    \centering
    \includegraphics[width=0.35\textwidth]{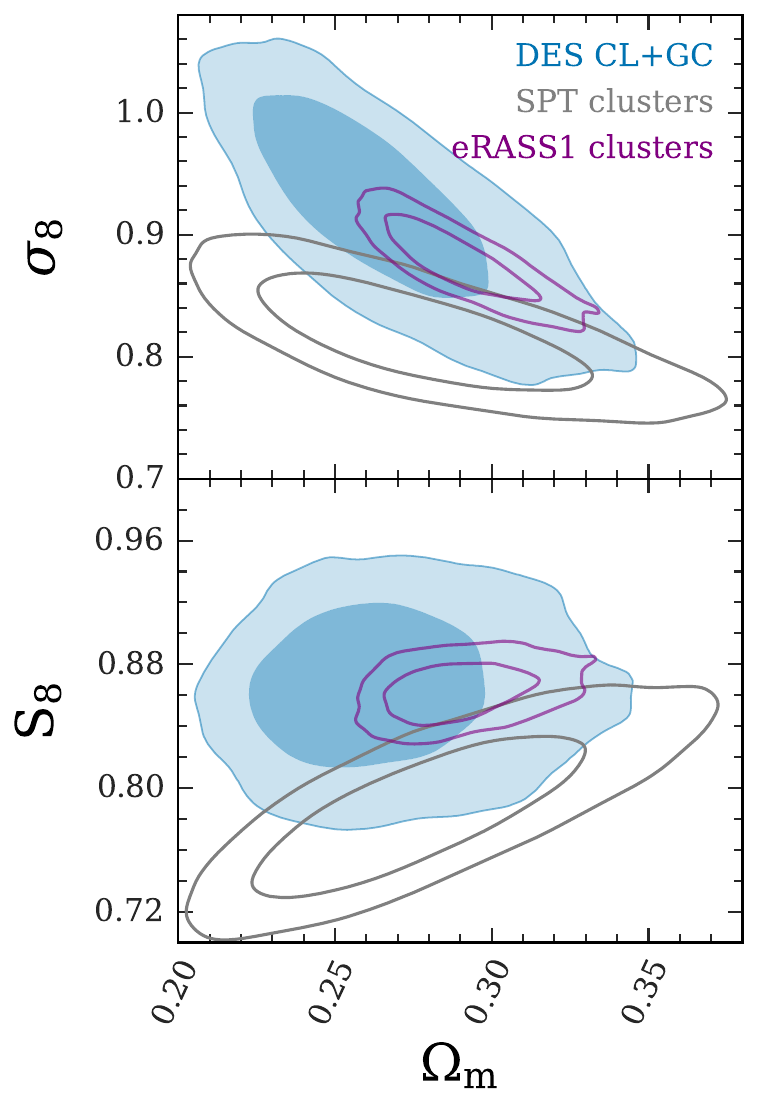}
    \caption{Marginalized constraints on the three key parameters $\sigma_8$, $S_8=\sigma_8\sqrt{\omegam/0.3}$, and $\omegam$ in the \nuCDM{} model from cluster cosmology analyses, including DES-Y3 clusters (blue), SPT-SZ+SPTpol clusters with DES-Y3 weak lensing mass calibrations (gray), and eRASS1 clusters with DES-Y3, HSC, and KiDS weak lensing mass calibrations (purple). Contours show $68\%$ and $95\%$ confidence intervals.}
    \label{fig:allcluster}
\end{figure}
\begin{figure}
    \centering
    \includegraphics[width=0.35\textwidth]{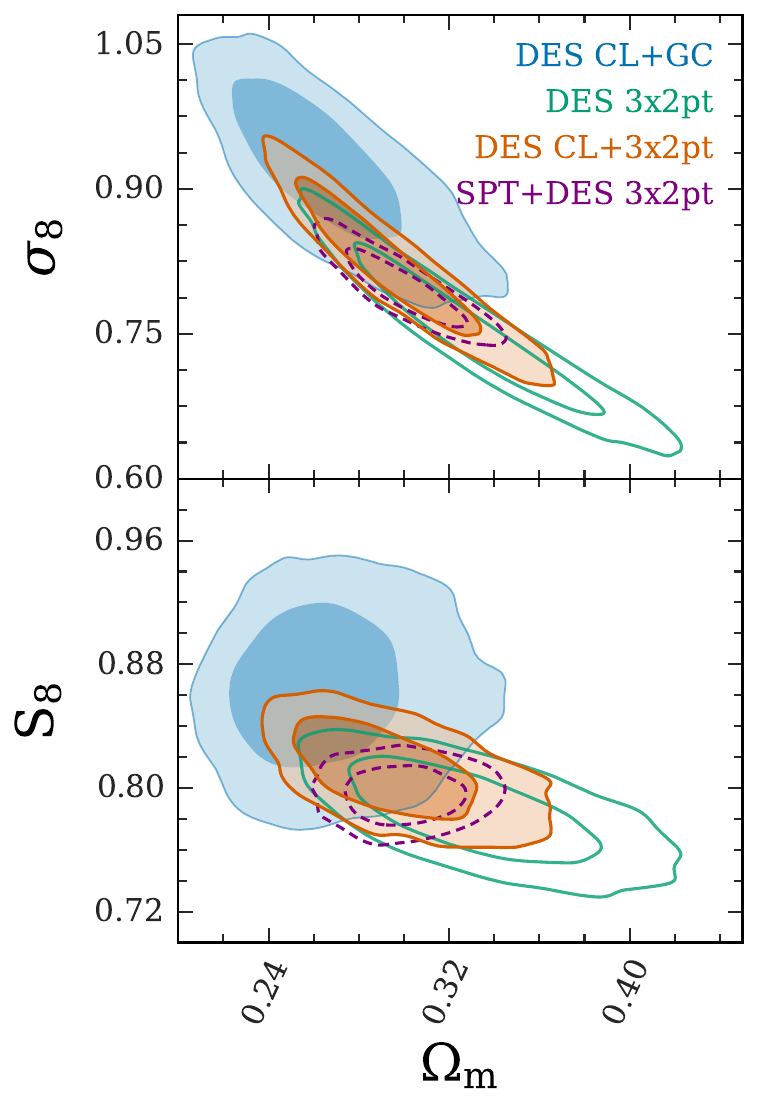}   
    \caption{Marginalized constraints on the three key parameters $\sigma_8$, $S_8=\sigma_8\sqrt{\omegam/0.3}$, and $\omegam$ in the \nuCDM{} model from different DES cosmological probes, including DES-Y3 clusters (blue), DES-Y3 \ttt{} (green), and joint analyses of DES-Y3 clusters and DES-Y3 \ttt{} (orange). We further show combined analyses of SPT clusters and DES-Y3 \ttt{} (purple). Contours show $68\%$ and $95\%$ confidence intervals.}
    \label{fig:lcdmDES}
\end{figure}

\begin{figure}
    \centering
    \includegraphics[width=0.35\textwidth]{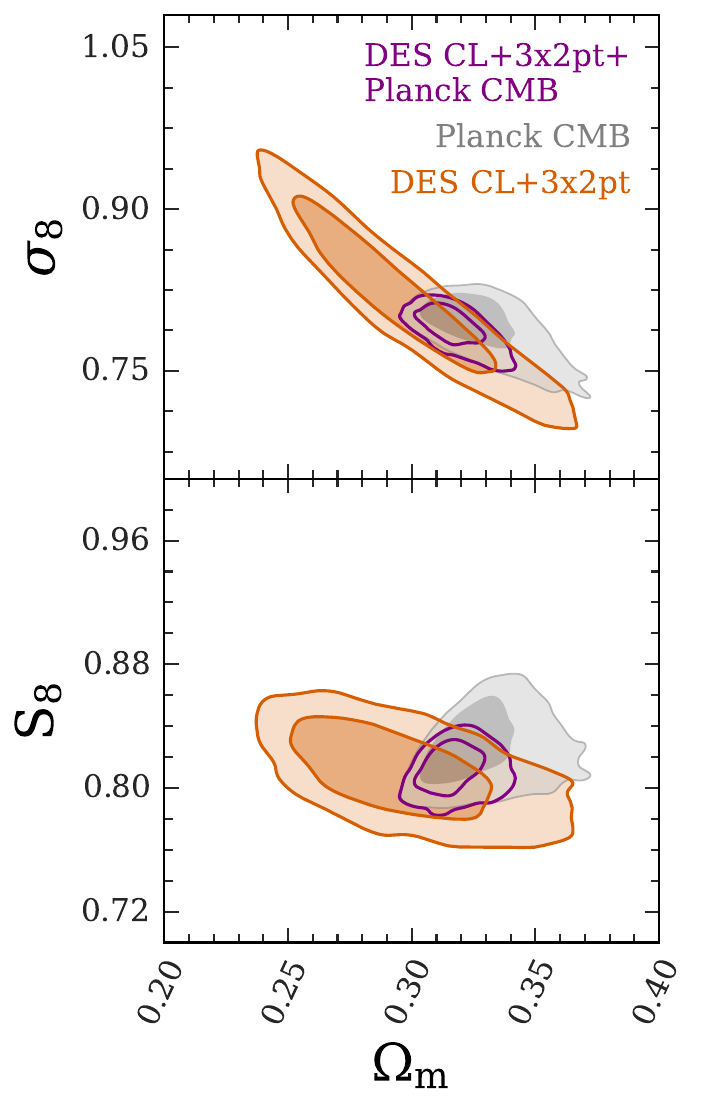}   
    \caption{Marginalized constraints on the three key parameters $\sigma_8$, $S_8=\sigma_8\sqrt{\omegam/0.3}$, and $\omegam$ in the \nuCDM{} model from the joint analysis of DES-Y3 clusters and DES-Y3 \ttt{} (orange). This measurement is further compared with the predictions from Planck CMB (gray). Given the consistency between different probes, we further show the combined constraints from DES \allcomb{} and Planck CMB (purple). Contours show $68\%$ and $95\%$ confidence intervals.}
    \label{fig:lcdmPlanck}
\end{figure}

\subsection{Consistency between \clustercomb{} cluster cosmology constraints and \ttt{} in DES-Y3}

We now turn to check the internal consistency between different DES cosmological probes. Using the PPD metric, we find $p(\xi_{\pm} | \rm{\clustercomb{}})=0.04$ and $p(\xi_{\pm}+\gamma_t [\rm{first\ three\  bin}] | \rm{\clustercomb{}}) = 0.07$. We note that we cannot calculate the PPD of full \ttt{} and \clustercomb{} because of shared galaxy clustering data vectors and the lack of the bias value of the highest redshift bin of \maglim{} in \clustercomb{} analyses. Finally, we check that the \nuCDM{} model fits to the combined data vector, obtaining $P(\rm{\clustercomb{}}|\nuCDM{})=0.53$. With all these tests, we established that DES-Y3 clusters and DES-Y3 \ttt{} are consistent under the \nuCDM{} model. This consistency itself is a remarkable cosmological test of the \nuCDM{} model because of the widely different masses and scales probed by different observables.

\subsection{Cosmology from joint analyses of cluster abundances, weak lensing, and galaxy clustering}
Having checked the consistency, we jointly analyze cluster abundances and all possible two-point correlation functions between cluster density, galaxy density, and weak lensing shear field, known as the \allcomb{} probes. The marginalized constraints on $\omegam$, $S_8$, and $\sigma_8$ are shown in Fig.~\ref{fig:lcdmDES} and summarized in Table~\ref{tab:paramsum}. Marginalized over $37$ astrophysical parameters, the DES \allcomb{} constraints on the key parameters are 
\begin{equation}
  \begin{aligned}
    S_8              &= 0.811^{+0.022}_{-0.020} \\
    \omegam          &= 0.294^{+0.022}_{-0.033} \\
    \sigma_8         &= 0.822\pm 0.053.
  \end{aligned}
\end{equation}
The ratio of the figure-of-merit on $\omegam$--$S_8$ of DES-Y3 \allcomb{} to DES-Y3 \ttt{} is $1.24$. In addition to the improved constraining power, \allcomb{} favors a higher $S_8$ and smaller $\omegam$ value than that of \ttt{}. In Fig.~\ref{fig:lcdmDES}, we also compare our \allcomb{} constraints with the combined analysis of SPT and DES-Y3 \ttt{} and find consistent results. The slightly better constraints of SPT and DES-Y3 \ttt{} are mostly due to a more orthogonal degeneracy direction between SPT and DES-Y3 \ttt{} than \clustercomb{} and DES-Y3 \ttt{}.

Comparing the \allcomb{}  with the prediction of \nuCDM{} based on Planck CMB TT, EE, TE likelihood re-analyzed with DES prior \citep{Y3kp}, we find that the parameter difference tension metric \citep{2021PhRvD.104d3504R} yields a PTE of $0.6$ ($0.85 \sigma$). The $S_8$ of \allcomb{} is $0.58\sigma$ lower than Planck under \nuCDM{} as shown in Fig.~\ref{fig:lcdmPlanck}.

\begin{figure}
    \centering
    \includegraphics[width=0.4\textwidth]{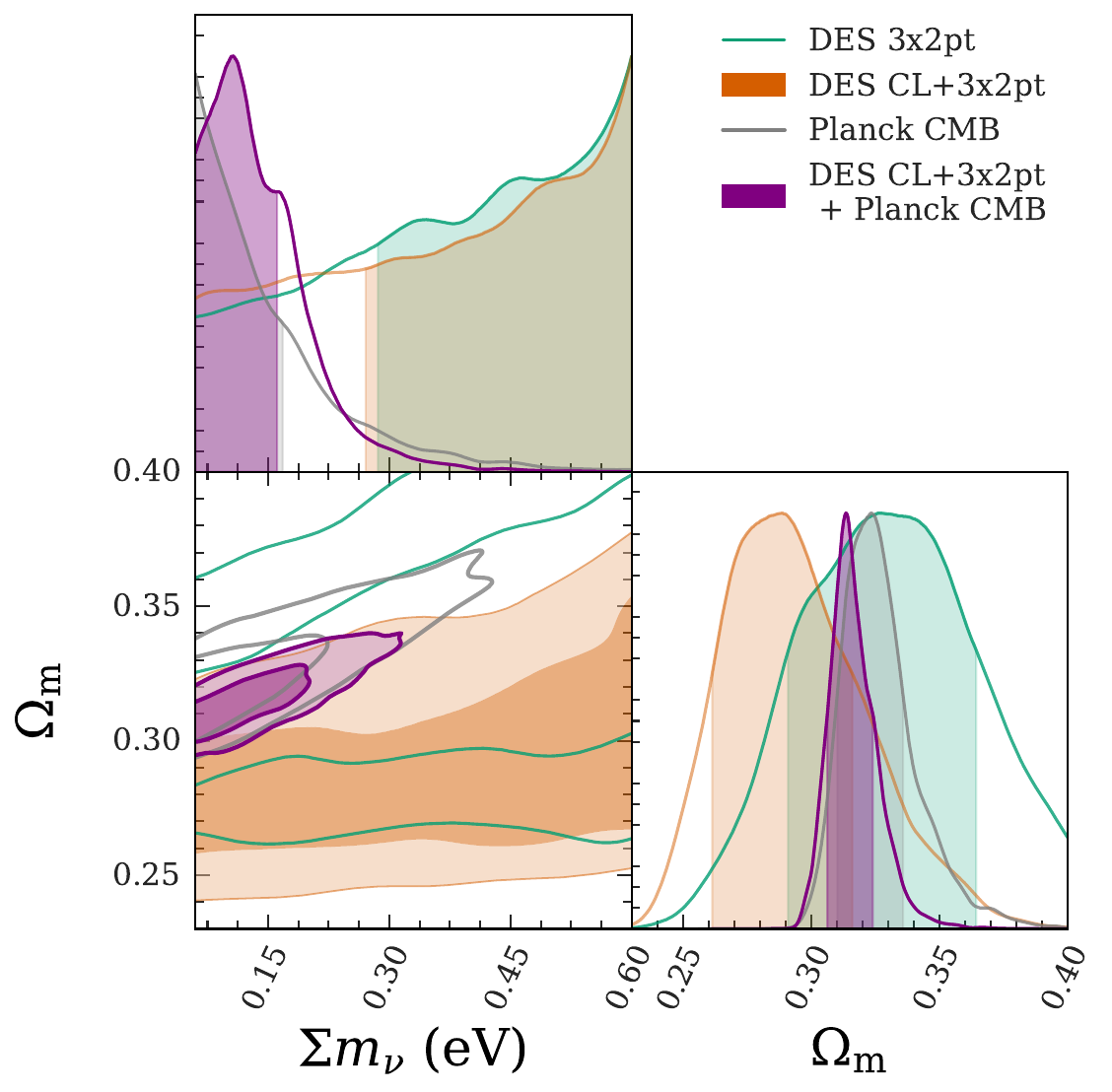}
    \caption{Marginalized constraints on the sum of neutrino masses $\sum m_\nu$ and $\omegam$ in the \nuCDM{} model. We show DES \ttt{} (green), DES \allcomb{} (orange), Planck (gray), and DES \allcomb{} and Planck CMB (purple). Contours show $68\%$ and $95\%$ confidence intervals. The upper panel shows marginalized posteriors for $\sum m_\nu$, with shaded regions showing the $68\%$ confidence interval. The right panel shows marginalized posteriors for $\omegam$, with shaded regions showing the $68\%$ confidence interval.} 
    \label{fig:lcmnu}
\end{figure}

Because DES-Y3 \allcomb{} and Planck CMB are consistent, we combine the two analyses to obtain tighter constraints on the cosmological parameters, which are summarized in Table~\ref{tab:paramsum}. In addition to the improved constraints on $S_8$ and $\omegam$, we show the constraints on the sum of neutrino masses in Fig.~\ref{fig:lcmnu}, where the neutrino mass and density $\Omega_\nu$ are related via $\sum m_\nu=93.14\Omega_\nu h^2$ eV. As shown in Fig.~\ref{fig:lcmnu}, the DES-Y3 \allcomb{} is able to break the degeneracy between $\omegam$ and $\sum m_\nu$ in the Planck-only constraint. Combining the DES \allcomb{} and Planck CMB leads to an upper limit 
\begin{equation}
    \sum m_\nu < 0.26 \rm{eV} \ (95\% \rm{CL}).
\end{equation}
This is a $\approx 65\%$ reduction compared to DES-Y3 \ttt{}+Planck \citep{Y3kp}, due to a greater constraining power of $\omegam$. Interestingly, the marginalized posterior of $\sum m_\nu$ peaks at $0.1$ eV, consistent with the combined constraints of SPT clusters and DES-Y3 \ttt{} \citep{SPTand3x2pt}.

\begin{figure}
    \centering
    \includegraphics[width=0.4\textwidth]{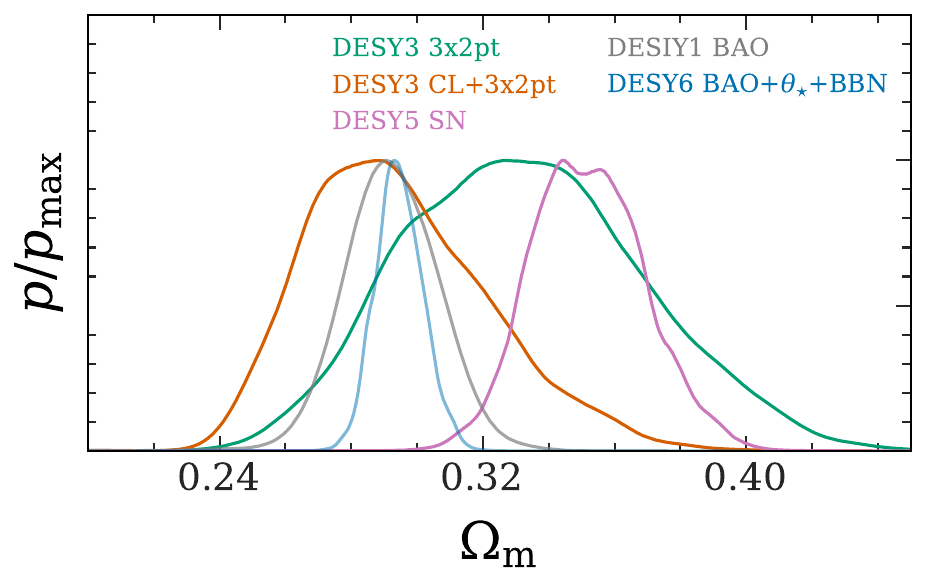}
    \caption{Marginalized posteriors on $\omegam$ in the \nuCDM{} model. We show DES \ttt{} (green), DES \allcomb{} (orange), DESI-Y1 BAO (gray, \citep{DESI_BAO}), DES-Y6 BAO+$\theta_\star$+BBN (blue, \citep{SNBAO}) and DES-Y5 supernovae (purple, \citep{DES_SN}). }
    \label{fig:omegam}
\end{figure}

Finally, we compare the $\omegam$ constraints from the DES-Y3 \allcomb{} analysis, DDES-Y3 SN \citep{DES_SN}, DES-Y6 BAO \citep{SNBAO}, and DESI-Y1 BAO constraints \citep{DESI_BAO} in Fig.~\ref{fig:omegam}. We find that the \allcomb{} constraints obtain a tighter constraint on $\omegam$ than \ttt{} and pull the value toward DESI BAO constraints. Compared to DES-Y5 SN, the \allcomb{} obtains an $\omegam$ value $2.04\sigma$ lower than that of DES-Y5 SN. This is a potentially intriguing deviation --  \citep{2024arXiv241204430T} showed that the differences between the $\omegam$ from DES-Y5 SN and DESI-Y1 BAO under $\nuCDM$ could be due to the evolution of dark energy equation of state. It would be interesting to investigate whether the difference between the $\omegam$ from DES-Y5 SN and DES-Y6 \allcomb{} is consistent with the prediction of the favored $w_0-w_a$ model in the joint analyses of DESI-Y1 BAO, DES-Y5 SN, and Planck CMB. However, our current model is not validated for the $w_0-w_a$ model; thus, we leave this investigation to future work. 

While we have been focusing on discussions of cosmological parameters, the \allcomb{} also provides a stringent constraint on the several nuisance parameters, which are presented in Appendix \ref{app:all}. In Appendix \ref{app:RMR}, we further show the inferred mass--richness relation of the DES-Y3 cluster samples based on the \allcomb{} analysis and detail the associated calculation.

\section{Conclusions}
\label{sec:conclusions}

This work presents the measurement, calibration, and cosmological constraints of cluster abundances and all possible two-point correlation functions between clusters, galaxies, and weak gravitational lensing shears measured in the first three years of DES data. Since DES-Y1 \citep{4x2pt2}, we have improved our analysis framework \citep{4x2pt1_duplicated} to meet the accuracy requirement of the much more constraining datasets, which covers about three times the sky area of that of DES-Y1. This improved analysis framework is described in detail and validated to meet the accuracy requirement of the full DES data in \citep{Y6clustermethod}. The computationally intensive analysis framework is enabled by a customized likelihood inference tool \citep{linna} that reduces the computation needs by a factor of $10$, making the computation consumptions of the project manageable. 

In the first phase of this work, we performed a blinded analysis on the cluster-based data vector (known as the \clustercomb{} analysis): the combination of cluster abundances, cluster lensing, cluster clustering, cluster--galaxy cross-correlation functions, and galaxy clustering. While carrying out the analysis, we shifted key cosmological and nuisance parameters by a random number. This allows us to test the robustness of our analysis and make decisions without knowing the actual cosmological parameters we would obtain. These decisions include the selection-effect model, survey systematics mitigation scheme, cluster lensing scale cuts and the analysis method, the tension metric, and the criteria for the goodness of fit. The blinding and unblinding processes are described in detail in Sec.~\ref{sec:blinding}. We do not alter any of the analysis after we unblind. 

With $\sim16,000$ optically detected clusters, our cluster-based cosmological constraint is the most powerful cosmological constraint from an optically selected cluster sample to date. We have achieved $\sim 50\%$ improvements in the constraints on the key cosmological parameters from our result in DES-Y1. We find that the $\nuCDM$ model is consistent with our cluster-based data vector with constraints on the clustering amplitude $S_8=0.864\pm 0.035$ and matter density $\omegam=0.265^{+0.019}_{-0.031}$. Comparing to cluster cosmological constraints in X-ray \citep{erasscluster} and SZ \citep{SPTcosmo}, we find that our cluster cosmological constraints are consistent with those analyses but have a slight preference for low $\omegam$ and high $S_8$. 

Under the $\nuCDM$ model, we find that our cluster cosmology constraint is consistent with the DES-Y3 \ttt{} constraints presented in \citep{Y3kp}. As clusters and DES-Y3 \ttt{} probe the universe in different environments and scales, the consistency of the constraints is itself a strong test of the $\nuCDM$ model. Given the consistency of the probes, we then perform a joint analysis of the DES-Y3 cluster and DES-Y3 \ttt{}, known as the \allcomb{} analysis. Marginalizing over $41$ nuisance parameters, we find constraints on the clustering amplitude $S_8=0.811^{+0.022}_{-0.020}$ and matter density $\omegam=0.294^{+0.022}_{-0.033}$. We have achieved $24\%$ improvements relative to DES-Y3 \ttt{} on the figure of merit on the $\omegam$--$S_8$ plane, similar to what we found in DES-Y1 \citep{4x2pt2}. Our $S_8$ constraint is $0.58\sigma$ lower than Planck under \nuCDM{}, which significantly weakens the claimed $S_8$ tension from some previous work, where the clustering amplitude measured by low-$z$ measurements tends to be $2-3\sigma$  lower than the prediction of Planck CMB under the $\nuCDM$ model. Our $\omegam$ is tighter than DES-Y3 \ttt{}, with the central value pulled toward DESI-BAO, and is $\sim 2\sigma$ lower than that of DES-Y5 SN. 

Given the strong consistency between our cosmological constraints and those from Planck CMB, we further combine DES-Y3 \allcomb{} and Planck CMB. We find a mild preference for a non-zero sum of neutrino masses with an upper limit $\sum m_\nu< 0.26$ eV [$95\%$ confidence interval] under the $\nuCDM$ cosmological model. This is consistent with the findings in the combined analyses of SPT, DES-Y3 \ttt{} and Planck CMB. 

This analysis presents the latest joint analyses of galaxy cluster abundances and \ttt{} in overlapping surveys, fully accounting for the cross-covariance between different cosmological probes. Our analysis framework has proven to reliably extract cosmological information from galaxy clusters detected in optical surveys and consistently deliver competitive constraints through the two stages
of the DES analysis \citep{4x2pt2}. We note that the analysis framework developed is not only useful for combining optical clusters with \ttt{} but also facilitates other combined analyses of clusters and \ttt{} \citep{SPTand3x2pt}. The novel advancement of the modeling and validation plan since DES-Y1 sets the foundation for future low-redshift multiprobe cosmological analyses in Stage-IV lensing surveys such as the Euclid mission, the Vera C. Rubin Observatory's Legacy Survey of Space and Time, and the Nancy Grace Roman Space Telescope.

\section{Acknowledgements.}

{\bf Author Contributions:}
All authors contributed to this paper and/or carried out infrastructure work that made this analysis possible. Some highlighted contributions from the authors of this paper include:
\textit{Scientific management and coordination}: Chun-Hao To and Matteo Costanzi (Cluster working group conveners).
\textit{Significant contributions to project development, including paper writing and figures}: Chihway Chang, Elisabeth Krause, Chun-Hao To, and Heidi Wu.
\textit{Data analysis and methods validation}: Elisabeth Krause, Eduardo Rozo, Chun-Hao To, Heidi Wu, and Risa Wechsler. 
\textit{Data vector generation}: Chun-Hao To. 
\textit{Internal reviewers}: Jonathan Blazek (Model and validation), Sebastian Bocquet (Model and validation), Agnès Ferté, Tesla E. Jeltema, and  Yuanyuan Zhang.
\textit{Advising}: David H. Weinberg. The remaining authors have made contributions to this paper that include, but are not limited to, the construction of DECam and other aspects of collecting the data; data processing and calibration; developing broadly used methods, codes, and simulations; running the pipelines and validation tests; and promoting the science analysis.

Funding for the DES Projects has been provided by the U.S. Department of Energy, the U.S. National Science Foundation, the Ministry of Science and Education of Spain, 
the Science and Technology Facilities Council of the United Kingdom, the Higher Education Funding Council for England, the National Center for Supercomputing 
Applications at the University of Illinois at Urbana-Champaign, the Kavli Institute of Cosmological Physics at the University of Chicago, 
the Center for Cosmology and Astro-Particle Physics at the Ohio State University,
the Mitchell Institute for Fundamental Physics and Astronomy at Texas A\&M University, Financiadora de Estudos e Projetos, 
Funda{\c c}{\~a}o Carlos Chagas Filho de Amparo {\`a} Pesquisa do Estado do Rio de Janeiro, Conselho Nacional de Desenvolvimento Cient{\'i}fico e Tecnol{\'o}gico and 
the Minist{\'e}rio da Ci{\^e}ncia, Tecnologia e Inova{\c c}{\~a}o, the Deutsche Forschungsgemeinschaft and the Collaborating Institutions in the Dark Energy Survey. 

The Collaborating Institutions are Argonne National Laboratory, the University of California at Santa Cruz, the University of Cambridge, Centro de Investigaciones Energ{\'e}ticas, 
Medioambientales y Tecnol{\'o}gicas-Madrid, the University of Chicago, University College London, the DES-Brazil Consortium, the University of Edinburgh, 
the Eidgen{\"o}ssische Technische Hochschule (ETH) Z{\"u}rich, 
Fermi National Accelerator Laboratory, the University of Illinois at Urbana-Champaign, the Institut de Ci{\`e}ncies de l'Espai (IEEC/CSIC), 
the Institut de F{\'i}sica d'Altes Energies, Lawrence Berkeley National Laboratory, the Ludwig-Maximilians Universit{\"a}t M{\"u}nchen and the associated Excellence Cluster Universe, 
the University of Michigan, NSF NOIRLab, the University of Nottingham, The Ohio State University, the University of Pennsylvania, the University of Portsmouth, 
SLAC National Accelerator Laboratory, Stanford University, the University of Sussex, Texas A\&M University, and the OzDES Membership Consortium.

Based in part on observations at NSF Cerro Tololo Inter-American Observatory at NSF NOIRLab (NOIRLab Prop. ID 2012B-0001; PI: J. Frieman), which is managed by the Association of Universities for Research in Astronomy (AURA) under a cooperative agreement with the National Science Foundation.

The DES data management system is supported by the National Science Foundation under Grant Numbers AST-1138766 and AST-1536171.
The DES participants from Spanish institutions are partially supported by MICINN under grants PID2021-123012, PID2021-128989 PID2022-141079, SEV-2016-0588, CEX2020-001058-M and CEX2020-001007-S, some of which include ERDF funds from the European Union. IFAE is partially funded by the CERCA program of the Generalitat de Catalunya.

We  acknowledge support from the Brazilian Instituto Nacional de Ci\^encia
e Tecnologia (INCT) do e-Universo (CNPq grant 465376/2014-2).

This document was prepared by the DES Collaboration using the resources of the Fermi National Accelerator Laboratory (Fermilab), a U.S. Department of Energy, Office of Science, Office of High Energy Physics HEP User Facility. Fermilab is managed by Fermi Forward Discovery Group, LLC, acting under Contract No. 89243024CSC000002.

We would like to thank Stanford University, the Stanford Research Computing Center, the Ohio Supercomputer Center,  and the University of Chicago’s Research Computing Center for providing the computational resources and support that contributed to these research results. 
\bibliographystyle{apsrev}
\bibliography{sample, refs_hw}
\clearpage
\onecolumngrid
\appendix

\section{Catalog update from \citep{Y3kp}}
\label{sec:catalog_update}

After the publication of \citep{Y3kp}, it was discovered that there was an inconsistency between the tomographic binning of the source catalog used for the data vector measurements and the redshift distribution used for the cosmological inference. The updated catalog has since been used by \citep{McCullough2024} %

\begin{figure}
    \centering
    \includegraphics[width=0.45\textwidth]{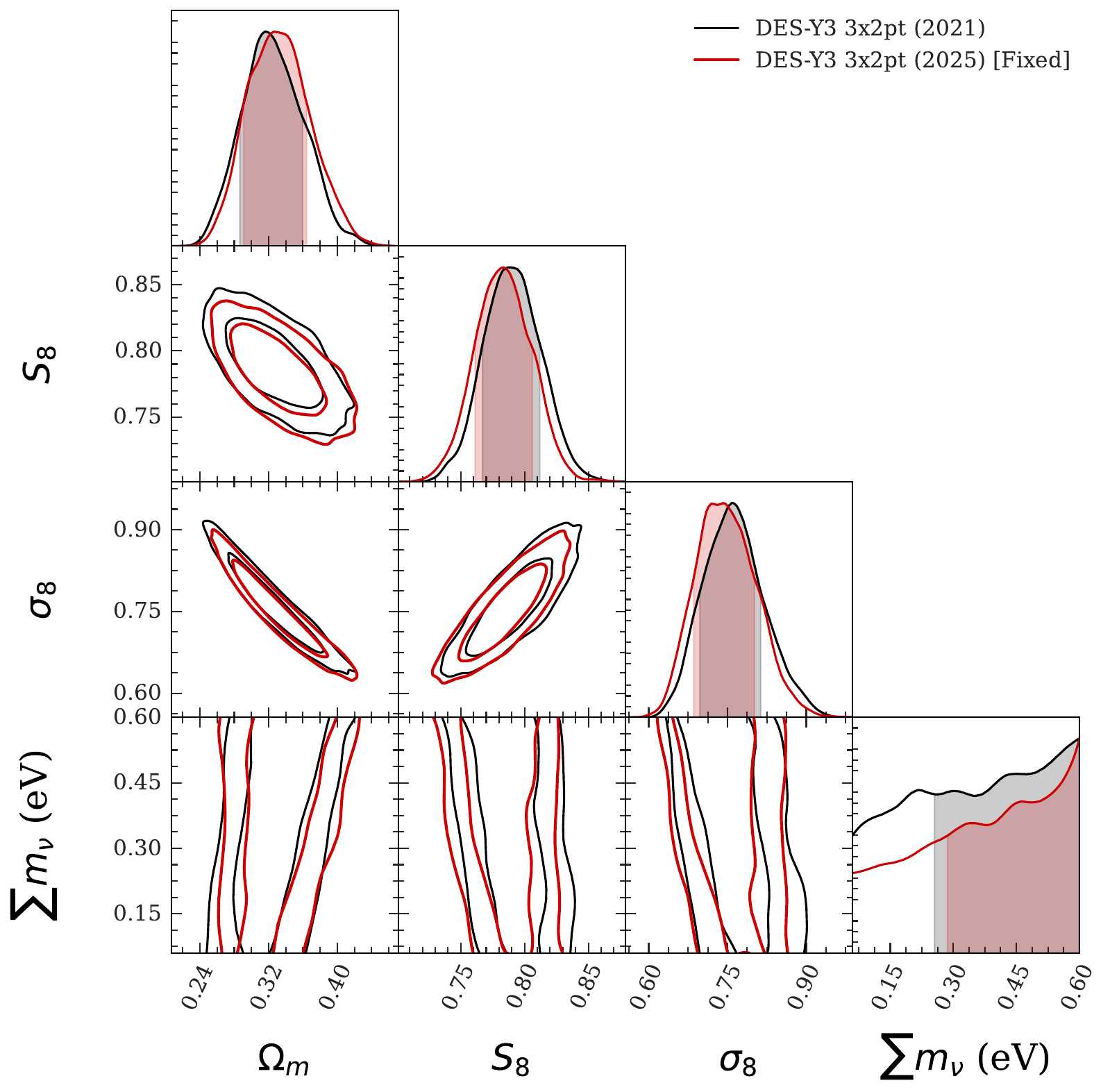}
        \includegraphics[width=0.45\textwidth]{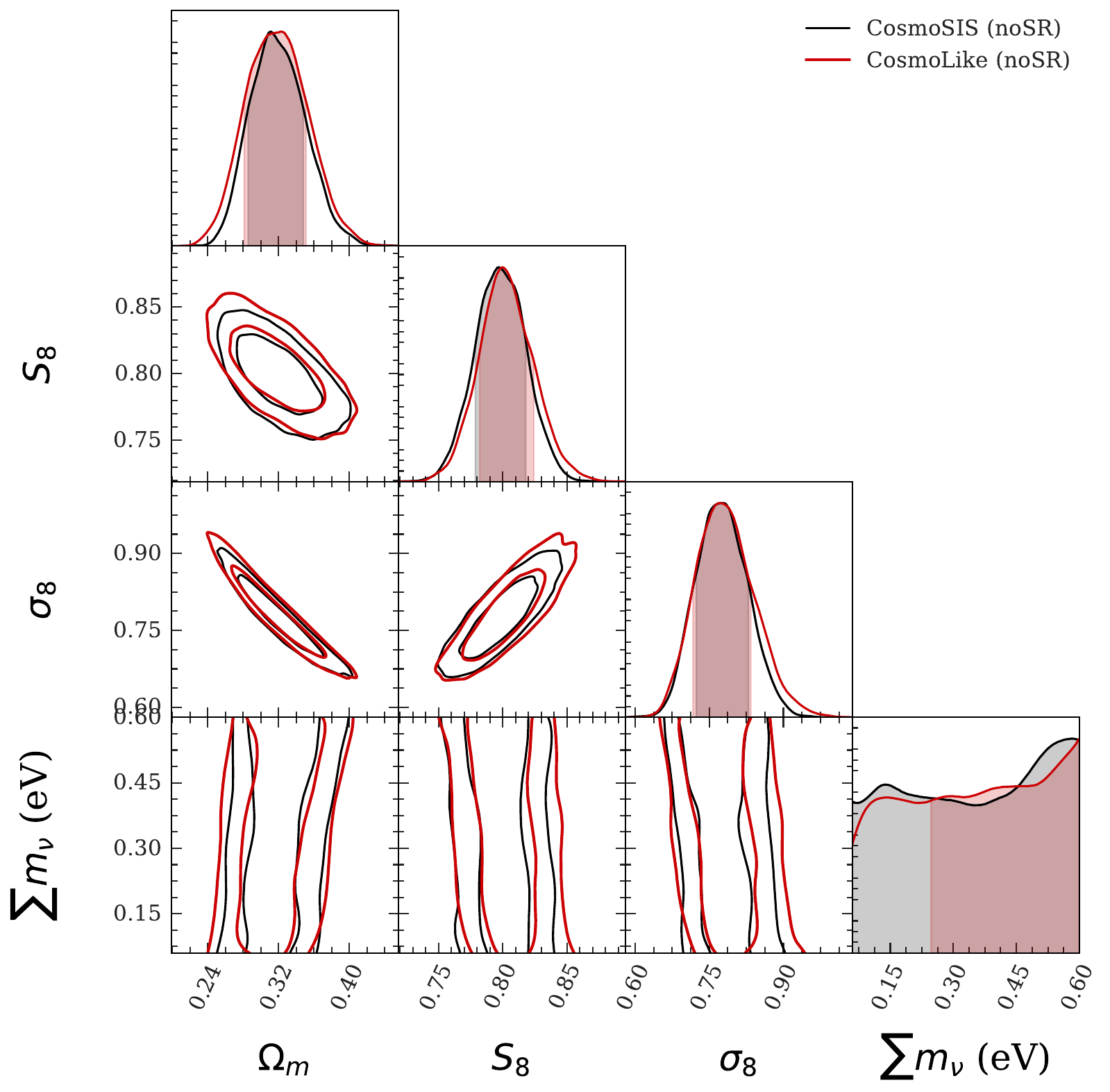}
    \caption{Left: Comparison of the 3$\times$2pt cosmological constraints from \citep{Y3kp} (black) and the same inference pipeline with the updated shear catalog (red). Right: Comparison of the updated 3$\times$2pt constraints using inference pipeline in \citep{Y3kp} and that used in this work, implemented through both \texttt{CosmoSIS} (black) and \texttt{CosmoLike} (red).}
    \label{fig:old_new_3x2}
\end{figure}

In this work, since we would like to combine the cluster probes with the 2$\times$2pt probes, we also present the updated cosmological constraints to \citep{Y3kp} using the corrected source catalog. Fig.~\ref{fig:old_new_3x2} compares the constraints from \citep{Y3kp} (black) and the updated data vectors running through the same analysis pipeline (red). We find that the updated constraints shift a negligible amount ($0.11 \sigma$ in $\omegam$ and $0.32 \sigma$ in $S_8$) from the published results, demonstrating that the cosmological constraints from \citep{Y3kp} remain robust. Interestingly, though perhaps expected, the new constraints also have a much better goodness-of-fit, going from $p$-value of $0.07$ to $0.47$. 

To facilitate the connection of this work with the published results in \citep{Y3kp}, in Fig.~\ref{fig:old_new_3x2}, we plot again the updated constraint and compare with the constraints using the same data vector but analysis pipeline adopted by this work (see Section~\ref{sec:model} and \citep{Y6clustermethod}) implemented both via \texttt{CosmoSIS} and \texttt{CosmoLike}. We find that \texttt{CosmoSIS} and \texttt{CosmoLike} give consistent cosmological constraints ($0.07\sigma$ in $\omegam$ and $0.2\sigma$ in $S_8$) while  \texttt{CosmoLike} is somewhat broader than that of  \texttt{CosmoSIS}. Given that the two analyses use sufficiently different samplers and modeling codes, this level of discrepancy is expected. We further compare the difference between \texttt{CosmoSIS} and \texttt{CosmoLike} predictions and find a difference of $\Delta \chi=0.20$ for \ttt{} and $\Delta \chi^2=0.06$ for 2$\times$2pt, similar to the findings in \citep{y3method}.

\section{Data vectors}
\label{sec:data_vec_plots}
We show in Fig.~\ref{fig:clusterN}, \ref{fig:gammac}, \ref{fig:wcc}, \ref{fig:wcg}, \ref{fig:wgg} the full set of data vectors used in this work. For all figures, each panel shows the data points in the upper part and the fractional difference between the data and the mean of the predictions from the \clustercomb{} chains, normalized by the PPD prediction scatter (Appendix~\ref{sec:ppd}). Data points excluded from the analysis by scale cuts are shown in light opacity.

\section{Posterior predictive distribution (PPD)}
\label{sec:ppd}
Here, we briefly summarize the PPD calculation. We refer the readers to \citep{y3-inttensions} for a more in-depth description and only focus on the differences in this section. Throughout the paper, we have used PPD for two purposes: 
\begin{enumerate}
    \item To quantify the goodness of fit between model ($\textbf{M}$) and data ($\textbf{d}$). 
    \item To quantify the consistency between two data vectors ($\textbf{d}_1$ and $\textbf{d}$), given a model ($\textbf{M}$). 
\end{enumerate}
To carry out these two calculations, we need to evaluate two probabilities $P(\textbf{d}|\textbf{M})$ and $P(\textbf{d}|\textbf{d}_1, \textbf{M})$ respectively. We approximate each probability as a Gaussian Mixture Model, written as 
\begin{eqnarray}
    P(\textbf{d}|\textbf{d}_1, \textbf{M})&=& \sum_i \phi^i \rm{Gauss}\left (\textbf{d}-\bm{\mu}_{2,i}-\textbf{C}^{-1}_{21}\textbf{C}^{-1}_{11}(\textbf{d}_1-\bm{\mu}_{1,i}), \textbf{C}_{22}-\textbf{C}_{21}\textbf{C}^{-1}_{11}\textbf{C}_{12} \right), \label{eq:ppd2} \\ 
    P(\textbf{d}|\textbf{M})&=& \sum_i \phi^i \rm{Gauss}(\textbf{d}-\bm{\mu}_i,\textbf{C}), \label{eq:ppd1}
\end{eqnarray}
where $i$ runs over the steps of the MCMC chains, $\phi^i$ are arbitrary normalization constants,  $\rm{Gauss(\textbf{x},\textbf{y})}$  denotes a multivariate Gaussian distribution with mean $\textbf{x}$ and covariance $\textbf{y}$, and $\textbf{C}$ is the covariance matrix of the data vector, $\textbf{C}_{\textbf{x}_1,\textbf{x}_2}$ is the covariance matrix between data vector $\textbf{x}_1$ and $\textbf{x}_2$. In the above expression, $\bm{\mu}_{i}$ is the theory prediction at step $i$ of the MCMC chain. For simplicity, we use $P(\textbf{d}, \textbf{M})$ to denote $P(\textbf{d}|\textbf{d}_1, \textbf{M})$ when testing the consistency of data and $P(\textbf{d}|\textbf{M})$  when testing the goodness-of-fit of the model.

To evaluate the consistency or the goodness of fit, we need to estimate whether the data at hand ($\textbf{d}_o$) is consistent with a random draw from the $P(\textbf{d},\textbf{M})$. We calculate the posterior predictive distribution (PPD) defined as 
\begin{eqnarray}
    \rm{PPD}(\textbf{d}_o|\textbf{M}) &:=& P\left(P(\textbf{d}_o|\textbf{M})>P(\textbf{d}_r|\textbf{M})\right) \\
    \rm{PPD}(\textbf{d}_o| \textbf{d}_1, \textbf{M}) &:=& P\left(P(\textbf{d}_o|\textbf{d}_1, \textbf{M})>P(\textbf{d}_r|\textbf{d}_1, \textbf{M})\right),
\end{eqnarray}
where $\textbf{d}_r$ is a random sample from $P(\textbf{d}|\textbf{M})$ in equation \ref{eq:ppd1} and from $P(\textbf{d}|\textbf{d}_1, \textbf{M})$ in equation \ref{eq:ppd2} . We numerically calculate the above probability with $15,000$ random draws from $P(\textbf{d}, \textbf{M})$. A low PPD value indicates that the data at hand is not a random draw from $P(\textbf{d}|\textbf{M})$, while a high PPD value could indicate a problem in the model, such as an overestimation of the covariance matrix. 

\begin{figure}
    \centering
\includegraphics[width=0.8\textwidth]{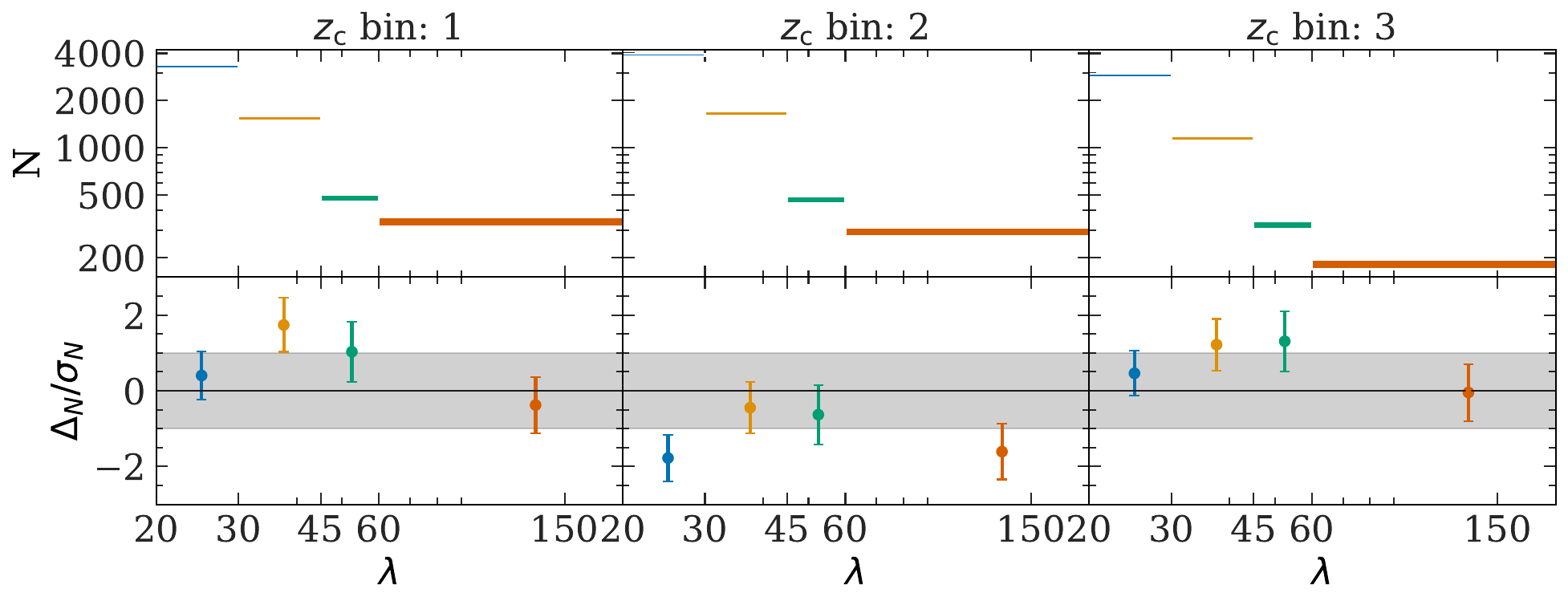}
    \caption{Measured cluster abundances for each tomographic bin. Each panel in column $i$ corresponds to measurements using \redmapper{} clusters in tomographic bin $i$. The shaded region represents $1\sigma$ uncertainties. The lower part of each panel shows fractional differences between the data and the mean prediction from the \clustercomb{} chains, normalized by the prediction scatter. Shaded bands denote the $1\sigma$ confidence interval.}
    \label{fig:clusterN}
\end{figure}

\begin{figure*}
    \centering
    \includegraphics[width=1.0\textwidth]{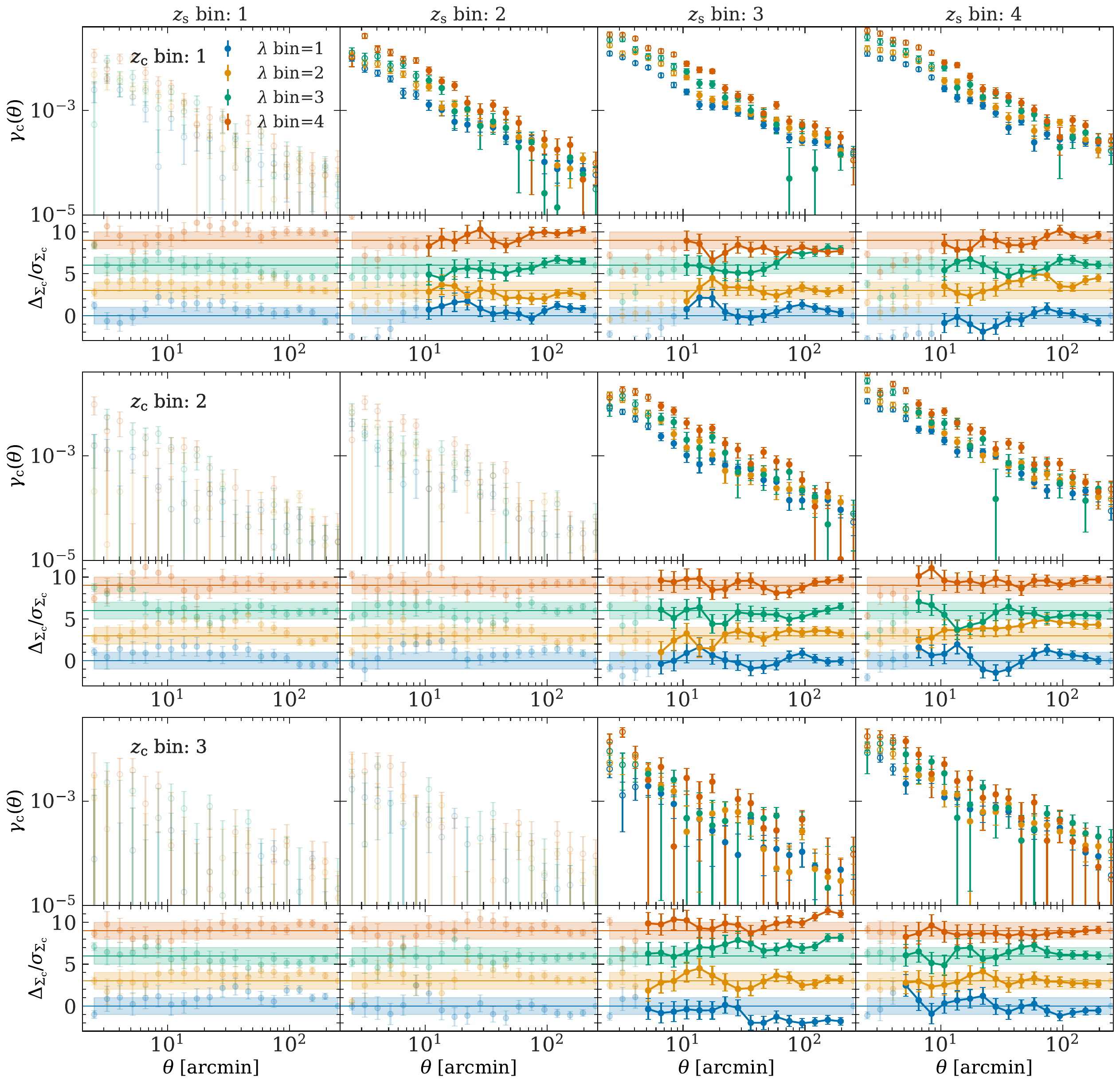}
    \caption{Measured $\gamma_c$ correlation functions for each tomographic bin combination. Each panel in row $i$ and column $j$ represents the measurement using clusters from tomographic bin $i$ and source galaxies from tomographic bin $j$. Colors indicate different richness bins, with error bars denoting $1\sigma$ uncertainties. Faint dots indicate data points excluded from the analysis. The lower part of each panel shows the fractional differences between the data and the mean of the predictions from the \clustercomb{} chains, normalized by the prediction scatter. For clarity, each richness bin is artificially shifted by 3. Shaded bands represent the $1\sigma$ confidence interval.}
    \label{fig:gammac}
\end{figure*}

\begin{figure*}
    \centering
    \includegraphics[width=1.0\textwidth]{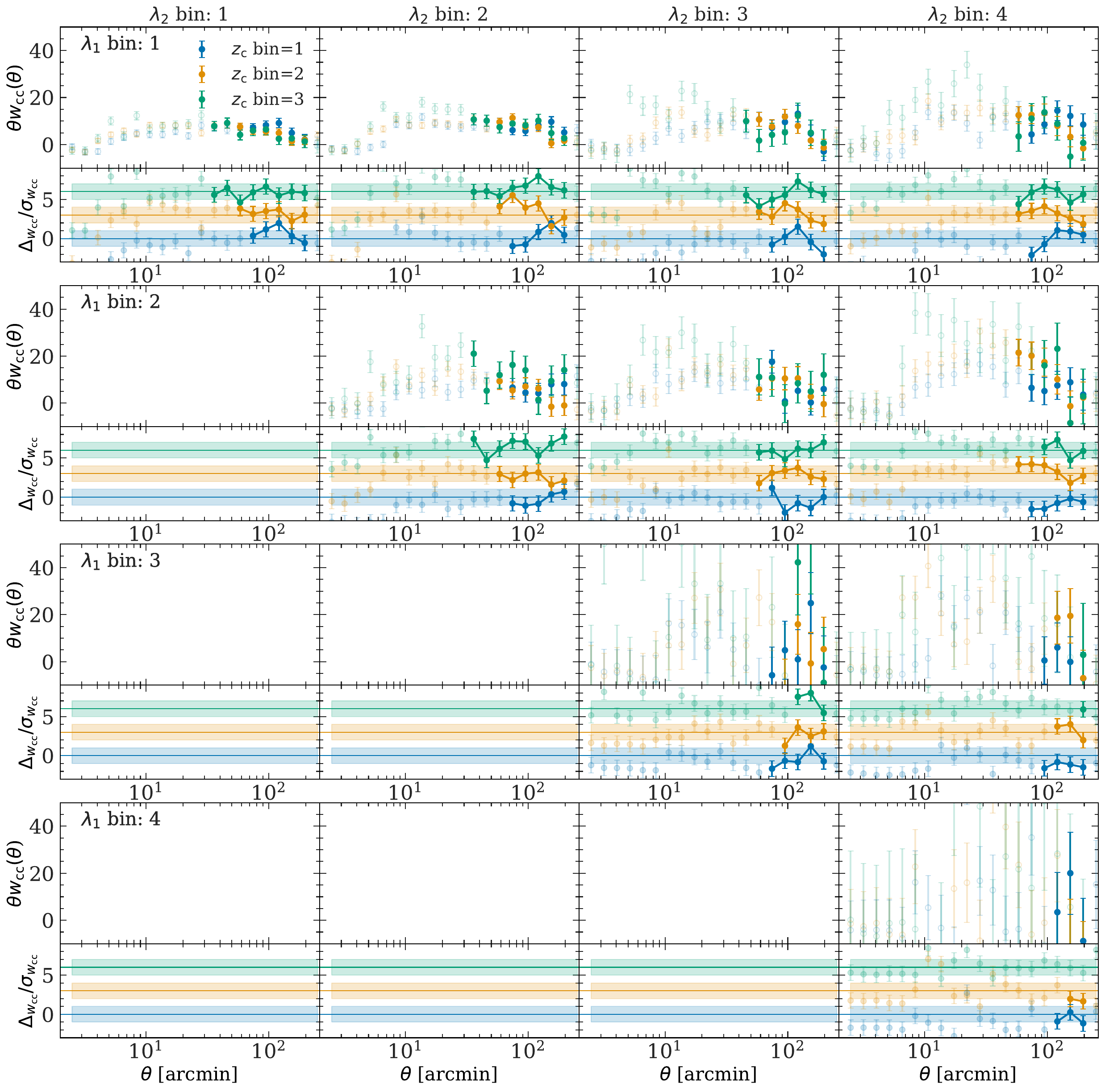}
    \caption{Measured $w_{\rm cc}$ correlation functions for each tomographic and richness bin combination. Each panel in row $i$ and column $j$ represents the measurement using clusters with richness $i$ and clusters with richness $j$. Colors indicate different redshift bins, with error bars denoting $1\sigma$ uncertainties. Faint dots indicate data points excluded from the analysis. The lower part of each panel shows the fractional differences between the data and the mean of the predictions from the \clustercomb{} chains, normalized by the prediction scatter. For clarity, each richness bin is artificially shifted by 3. Shaded bands represent the $1\sigma$ confidence interval.}
    \label{fig:wcc}
\end{figure*}

\begin{figure*}
    \centering
    \includegraphics[width=1.0\textwidth]{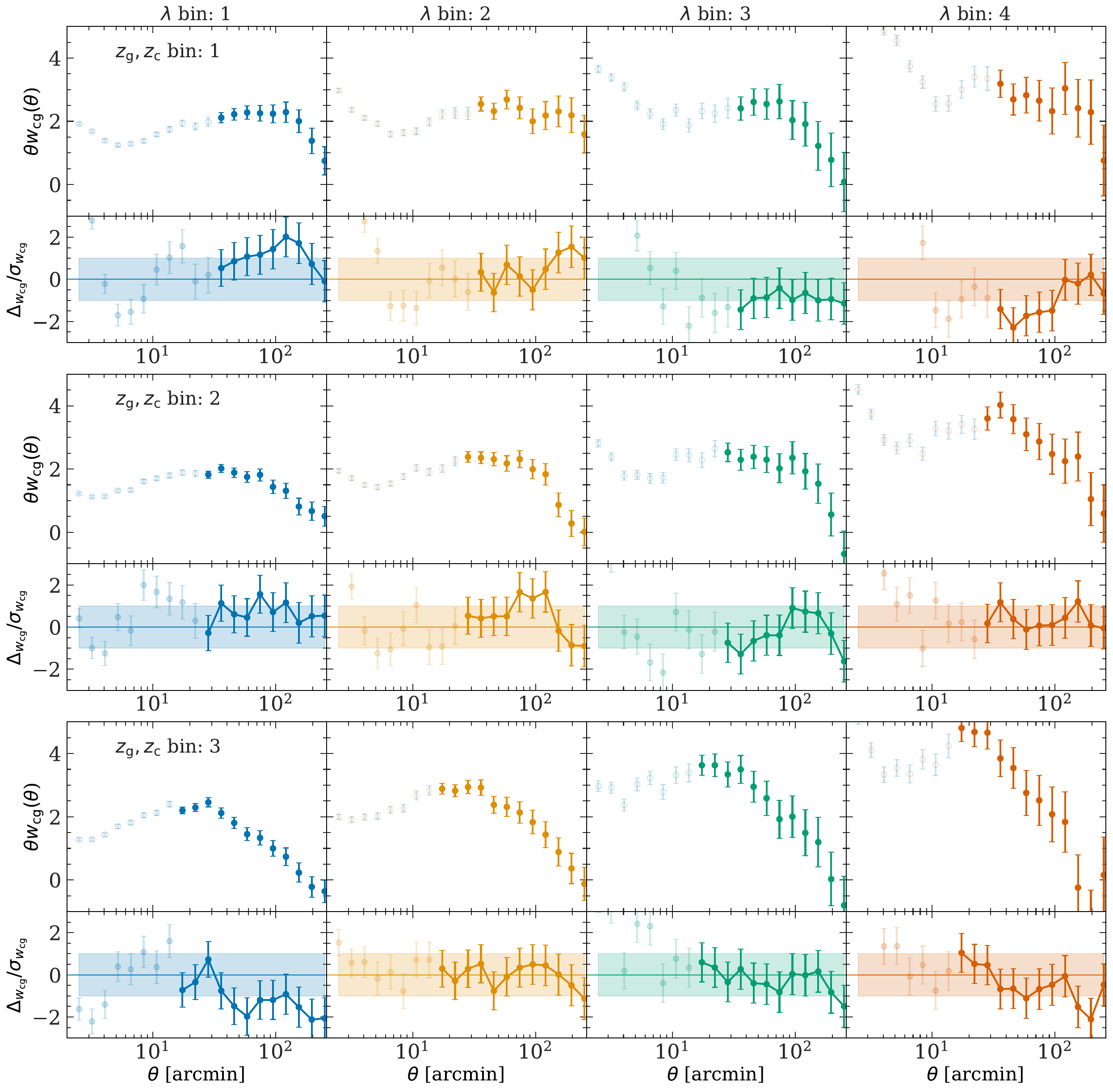}
    \caption{Measured $w_{\rm cg}$ correlation functions for each tomographic and richness bin combination. Each panel in row $i$ and column $j$ represents the measurement using clusters from tomographic bin $i$ with richness $j$ and \maglim{} galaxies from tomographic bin $i$. Colors indicate different richness bins, with error bars denoting $1\sigma$ uncertainties. Faint dots indicate data points excluded from the analysis. The lower part of each panel shows the fractional differences between the data and the mean of the predictions from the \clustercomb{} chains, normalized by the prediction scatter. For clarity, each richness bin is artificially shifted by 3. Shaded bands represent the $1\sigma$ confidence interval.}
    \label{fig:wcg}
\end{figure*}

\begin{figure}
    \centering
    \includegraphics[width=1.0\textwidth]{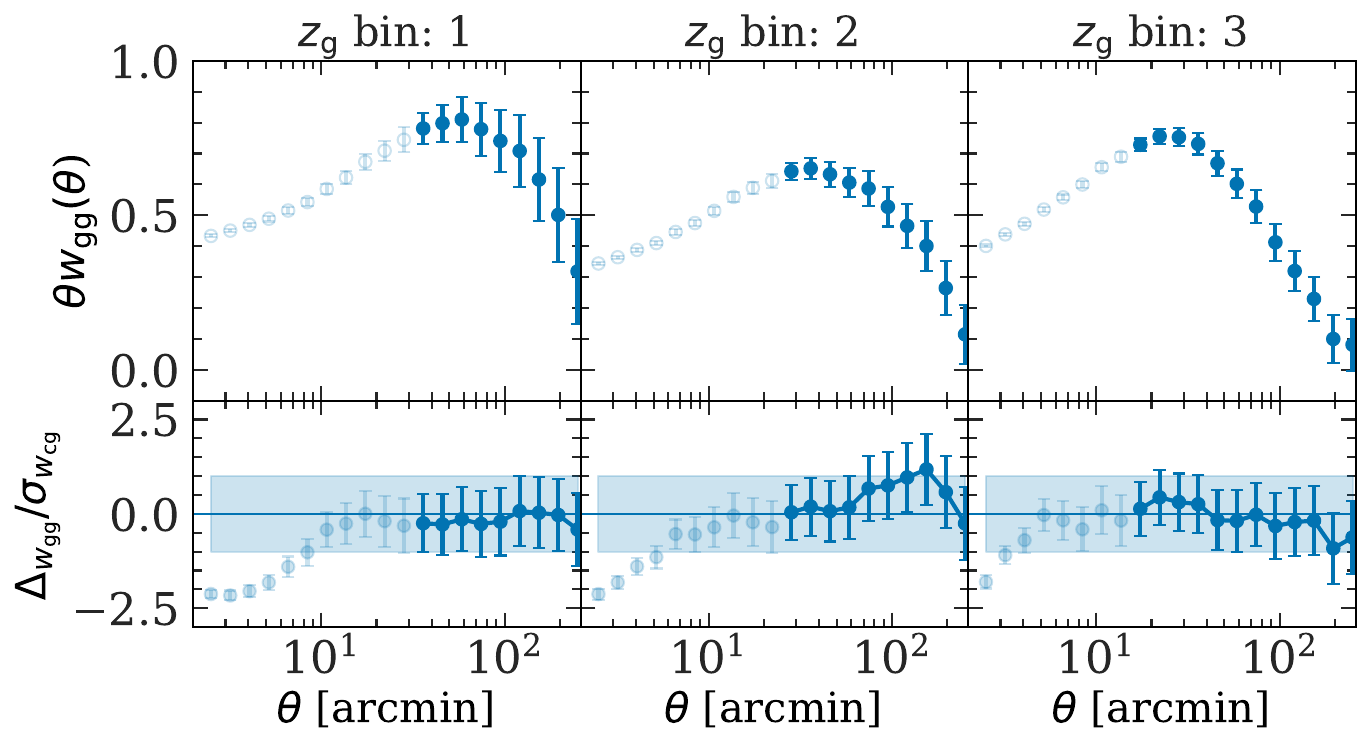}
    \caption{Measured $w_{\rm gg}$ correlation functions for each tomographic bin. Each panel in column $i$ corresponds to measurements using \maglim{} galaxies in tomographic bin $i$. Error bars represent $1\sigma$ uncertainties, with faint dots indicating data points excluded from the analysis. The lower part of each panel shows fractional differences between the data and the mean prediction from the \clustercomb{} chains, normalized by the prediction scatter. For clarity, each richness bin is offset by 3. Shaded bands denote the $1\sigma$ confidence interval.
}
    \label{fig:wgg}
\end{figure}

Finally, with a large number of draws, $P(\textbf{d}_r)$ can be approximated as a Gaussian distribution. We can evaluate the mean and the standard deviation from the $P(\textbf{d}_r)$ and compare it with the data vector at hand $\textbf{d}_o$. While less accurate, this comparison can be used as a visual check on whether the data and model are compatible. We plot this comparison in the lower panel of Fig.~\ref{fig:clusterN}, \ref{fig:gammac}, \ref{fig:wcc}, \ref{fig:wcg}, \ref{fig:wgg} and do not find any obvious deviation. 

\section{Mass--observable relations}
\label{app:RMR}
\begin{figure}
    \centering
    \includegraphics[width=1.0\textwidth]{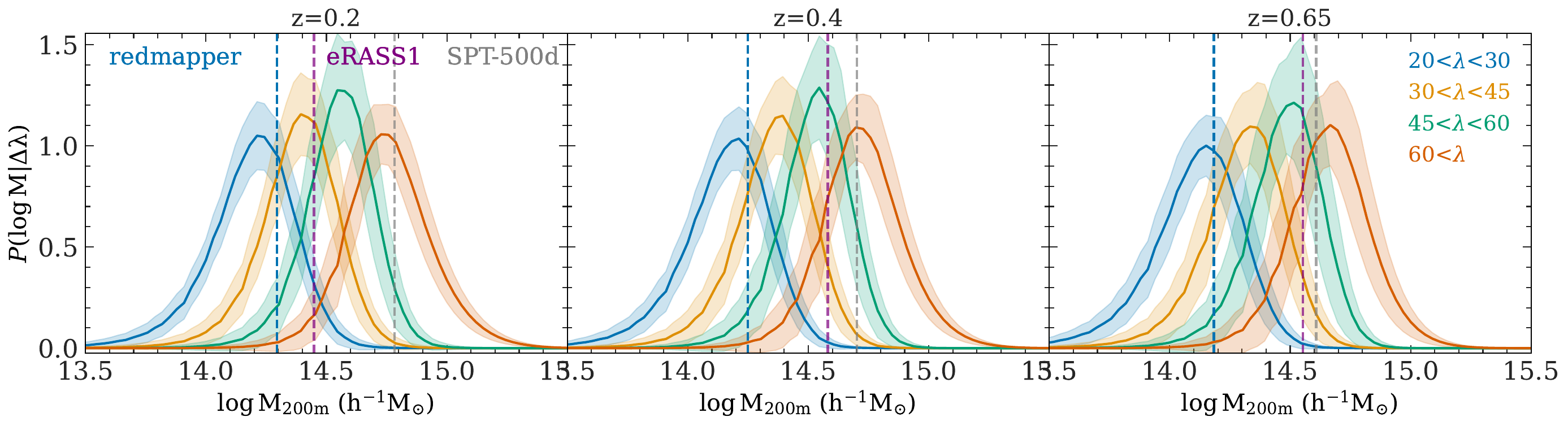}
    \caption{Distribution of \redmapper{} cluster mass in each richness bin. The width of the bands corresponds to the $68\%$ confidence interval of the distribution sampled from the posterior of the \allcomb{} analysis. Dashed lines show median mass of the samples in each redshift, including  \redmapper{} (blue), eRASS1 (purple), SPTpol (gray).
}
    \label{fig:MOR}
\end{figure}

We calculate the mass distribution of our samples using the posterior from the \allcomb{} analysis. Specifically, the mass distribution of a cluster given a richness bin $\Delta \lambda_c$ at redshift $z$  can be calculated as 
\begin{eqnarray}
    P(M|\Delta \lambda, z) = \frac{\int_{\lambda\in \Delta \lambda_c} n(M,z) P(\lambda|M,z) d\lambda}{\int_{\lambda\in \Delta \lambda} n(M,z)  d\lambda}, 
\end{eqnarray}
where $n(M,z)$ is the halo mass function and $P(\lambda|M, z)$ is the richness--mass relation. 
The mass distribution of \redmapper{} is shown in Fig.~\ref{fig:MOR}. We further show comparison with SPT-Pol \citep{2024OJAp....7E..13B} and eRASS1 \citep{erasscluster}. To facilitate the comparison, we use the \textsc{colossus} \citep{2018ApJS..239...35D} package to convert $M_{500\rm{c}}$ to $M_{200\rm{m}}$ and  assuming an NFW profile with a concentraion-mass relation \citep{2019ApJ...871..168D}.

Further, the \allcomb{} leads to a stringent constraint on the mean halo mass--richness relation. Marginalized over cosmological and nuisance parameters, the mean mass of \redmapper\ clusters is constrained as
\begin{eqnarray}
    \langle M_{200\rm{m}} | \lambda \rangle = 10^{14.399 \pm 0.011}\left(\frac{\lambda}{40}\right)^{1.053\pm 0.031} \left(\frac{1+z_\lambda}{1.45}\right)^{-0.667\pm 0.194} \nonumber h^{-1}M_{\odot}.
\end{eqnarray}
Since the richness changes between DES-Y1 and DES-Y3 (Fig.~\ref{fig:redmappercomparison}), it is hard to compare this value with existing literature. However, we note that while the normalization changes, we find that the constrained slope of the mass--richness relation is consistent with those in the literature \citep{4x2pt2, Costanzi2021}.

\section{All parameters}
\label{app:all}
\begin{figure}
    \centering
    \includegraphics[width=1.0\textwidth]{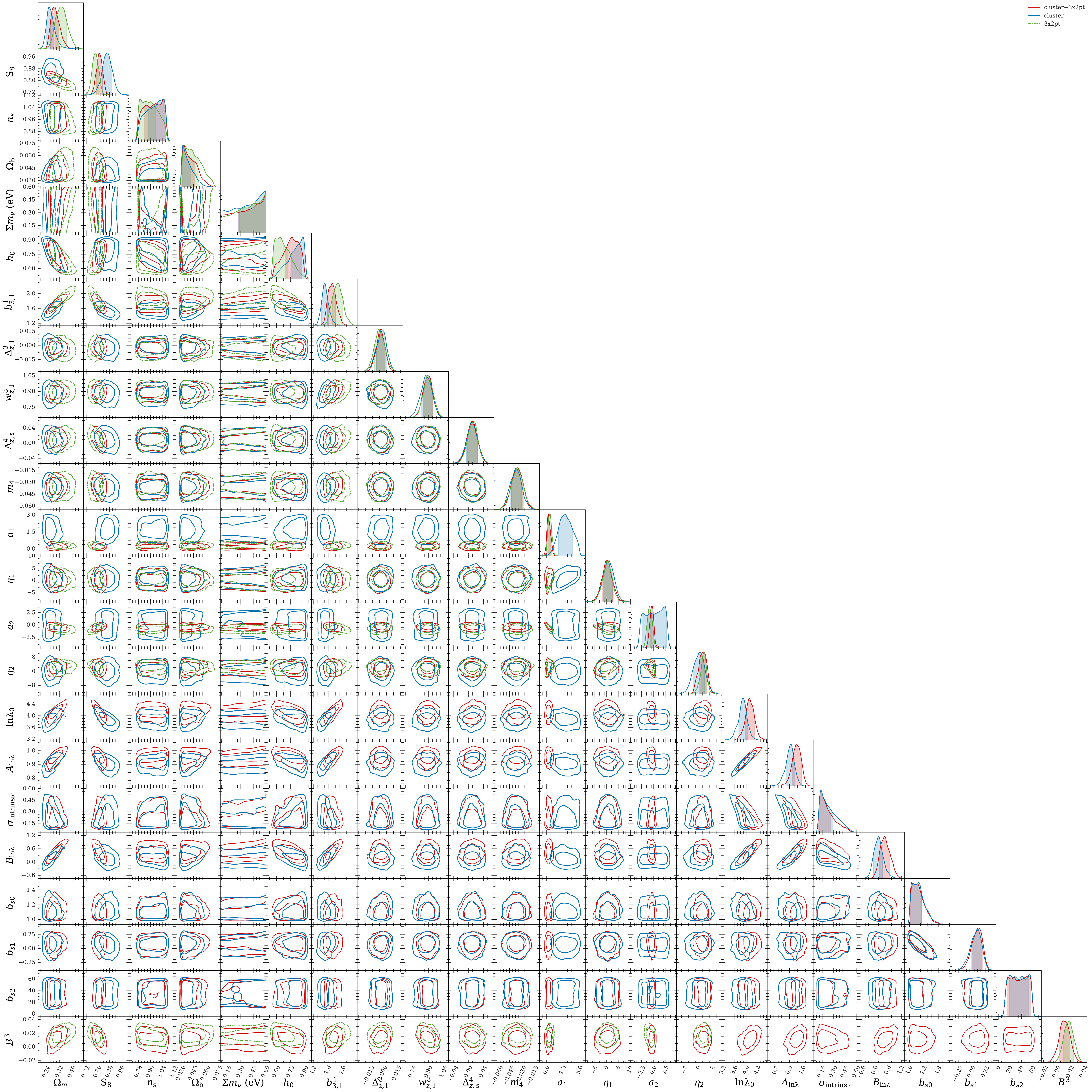}
    \caption{Summary cosmology and selected nuisance parameters for \clustercomb{} (blue), \ttt{} (green), and \allcomb{} (red). Contours show $68\%$ and $95\%$ confidence intervals. For lens galaxies, we only show parameters related to the third \maglim{} bin, which is the highest redshift bin for \clustercomb{} analysis. For source galaxies, we show parameters related to the fourth \mcal{} bin. 
}
    \label{fig:allp}
\end{figure}

We show constraints of nuisance parameters in Fig.~\ref{fig:allp}. The one-dimensional mean and $1\sigma$ confidence intervals of the nuisance parameters are summarized in Table~ \ref{tab:nuisance_param}. Interestingly, we find that the selection bias is consistent with $1$, indicating no detection of selection effect on large scales. This is consistent with our findings in DES-Y1 \citep{4x2pt2}.

\begin{table*}
    \centering
    \footnotesize
    \caption{Summary of the mean and $1\sigma$ confidence interval of the nuisance parameters constrained by \clustercomb{} and \allcomb{}. Parameters that are not constrained are indicated by a dash.}
    \label{tab:nuisance_param}
   \begin{tabular}{ lcc}
    \hline 
	Parameter	 & \allcomb{} & \clustercomb{} \\
	\hline 
	\textbf{Galaxy Bias} && \\
	$b_{1,\rm{l}}^1$ & $1.194^{+0.068}_{-0.094}   $ &$1.356^{+0.083}_{-0.100}    $\\ 
$b_{1,\rm{l}}^2$ & $1.494^{+0.081}_{-0.120}    $ &$1.562^{+0.090}_{-0.120}    $\\ 
$b_{1,\rm{l}}^3$ & $1.519^{+0.085}_{-0.110}    $ &$1.709^{+0.093}_{-0.130}    $\\ 
$b_{1,\rm{l}}^4$ & - &$1.630^{+0.100}_{-0.120}      $\\ 
	\hline
 \textbf{Intrinsic Alignment}&& \\
 $a_1$ &$1.717\pm 0.604$ &$0.203^{+0.143}_{-0.189}    $\\
  $\eta_1$ &$0.735\pm 2.346    $ &$0.255\pm 2.276              $ \\
  $a_2$ & -&$-0.403^{+0.435}_{-0.374}    $ \\
  $\eta_2$ &- & $2.020^{+2.640}_{-2.089}       $\\
 \hline
\textbf{Point-mass Marginalization}&&\\
	$B^0$ & -&$0.008^{+0.007}_{-0.007}$\\
    $B^1$ & -&$-0.001\pm 0.008         $\\
    $B^2$ &- &$0.011\pm 0.009         $\\
    $B^3$ & -&$0.006\pm 0.010            $\\
	\hline
    \textbf{\redmapper{} Richness--Mass Relation} && \\
    $\rm{ln}\lambda_{0}$ & $3.874^{+0.167}_{-0.136}$ &$4.122^{+0.162}_{-0.146}     $\\
    $A_{\rm{ln}\lambda}$ & $0.903^{+0.037}_{-0.029}$ &$0.953^{+0.037}_{-0.030}   $\\
    $B_{\rm{ln}\lambda}$ & $0.163^{+0.183}_{-0.245} $ &$0.462^{+0.209}_{-0.263}     $\\
    $\sigma_{\rm{intrinsic}}$ & $0.238^{+0.053}_{-0.140}$ &$0.222^{+0.041}_{-0.120}    $\\
    \hline 
    \textbf{\redmapper{} Selection Effect} && \\
    $b_{s1}$ &$1.128^{+0.044}_{-0.110}$  &$1.123^{+0.043}_{-0.110}    $\\
    $b_{s2}$ & $0.067^{+0.120}_{-0.089}    $ &$0.086^{+0.110}_{-0.084}    $\\
    $r_0$ & - & -\\
    \hline
    \end{tabular}
\end{table*}

\section{Comparisons with previous DES cluster analyses}
\label{app:descompare}
\begin{figure}
    \centering
    \includegraphics[width=0.5\textwidth]{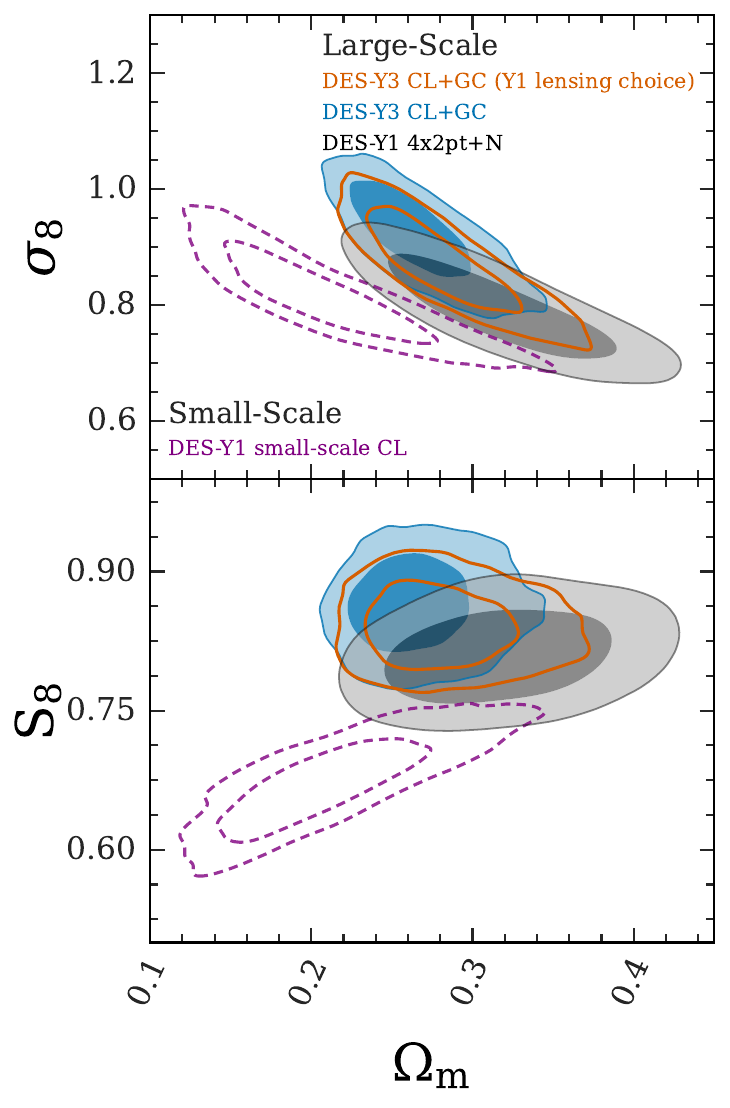}
    \caption{Comparison of DES cluster cosmology analyses, including DES-Y3 \clustercomb{} analysis (blue), DES-Y3 \clustercomb{} analysis with Y1 lensing analyses choices (orange), DES-Y1 \clustercomb{} analysis (gray, \citep{4x2pt2}). We further compare with the analysis based on small-scale cluster lensing and cluster abundances (purple, \citep{2023arXiv230906593A}).
}
    \label{fig:desy1comp}
\end{figure}

Figure~\ref{fig:desy1comp} compares DES  cluster cosmology analyses. The comparison of the large-scale-based analysis between DES-Y1 and DES-Y3 was discussed in section \ref{section:unblind}. Specifically, the analysis of Y3 data with Y1 cluster lensing analysis choices shifts the contour toward DES-Y1 constraints. However, we believe that Y1 cluster lensing analysis has residual contaminations from small-scale cluster lensing due to the lensing estimator being non-local. We decided to adopt the Y3 analysis choice as fiducial before we unblind the parameter constraints. As a comparison, we also show constraints from DES-Y1 cluster analyses using small-scale cluster lensing information, presented in \citep{2023arXiv230906593A}.
These constraints differ slightly from the fiducial constraints presented in \citep{DES_cluster_cosmology}, with a 0.5$\sigma$ shift in the $\omegam$–$\sigma_8$ plane, which is due to differences in the sampling methods and the adopted richness--mass relations. 
The DES-Y1 cluster analyses using small-scale cluster lensing are known to be affected by selection effects \citep{DES_cluster_cosmology, Costanzi2021} and are inconsistent with the large-scale analyses.

\textsuperscript{1} Cerro Tololo Inter-American Observatory, NSF's National Optical-Infrared Astronomy Research Laboratory, Casilla 603, La Serena, Chile\\
\textsuperscript{2} Laborat\'orio Interinstitucional de e-Astronomia - LIneA, Av. Pastor Martin Luther King Jr, 126 Del Castilho, Nova Am\'erica Offices, Torre 3000/sala 817 CEP: 20765-000, Brazil\\
\textsuperscript{3} Institute of Space Sciences (ICE, CSIC),  Campus UAB, Carrer de Can Magrans, s/n,  08193 Barcelona, Spain\\
\textsuperscript{4} Kavli Institute for Cosmological Physics, University of Chicago, Chicago, IL 60637, USA\\
\textsuperscript{5} Physik-Institut, University of Zürich, Winterthurerstrasse 190, CH-8057 Zürich, Switzerland\\
\textsuperscript{6} Centro de Investigaciones Energ\'eticas, Medioambientales y Tecnol\'ogicas (CIEMAT), Madrid, Spain\\
\textsuperscript{7} Institute of Cosmology and Gravitation, University of Portsmouth, Portsmouth, PO1 3FX, UK\\
\textsuperscript{8} Argonne National Laboratory, 9700 South Cass Avenue, Lemont, IL 60439, USA\\
\textsuperscript{9} Department of Physics and Astronomy, Pevensey Building, University of Sussex, Brighton, BN1 9QH, UK\\
\textsuperscript{10} Department of Physics, Northeastern University, Boston, MA 02115, USA\\
\textsuperscript{11} University Observatory, Faculty of Physics, Ludwig-Maximilians-Universit\"at, Scheinerstr. 1, 81679 Munich, Germany\\
\textsuperscript{12} Department of Physics \& Astronomy, University College London, Gower Street, London, WC1E 6BT, UK\\
\textsuperscript{13} Instituto de Astrofisica de Canarias, E-38205 La Laguna, Tenerife, Spain\\
\textsuperscript{14} Universidad de La Laguna, Dpto. Astrofísica, E-38206 La Laguna, Tenerife, Spain\\
\textsuperscript{15} Institut de F\'\\
\textsuperscript{16} Institut d'Estudis Espacials de Catalunya (IEEC), 08034 Barcelona, Spain\\
\textsuperscript{17} Department of Astronomy and Astrophysics, University of Chicago, Chicago, IL 60637, USA\\
\textsuperscript{18} NASA Goddard Space Flight Center, 8800 Greenbelt Rd, Greenbelt, MD 20771, USA\\
\textsuperscript{19} Jodrell Bank Center for Astrophysics, School of Physics and Astronomy, University of Manchester, Oxford Road, Manchester, M13 9PL, UK\\
\textsuperscript{20} University of Nottingham, School of Physics and Astronomy, Nottingham NG7 2RD, UK\\
\textsuperscript{21} Astronomy Unit, Department of Physics, University of Trieste, via Tiepolo 11, I-34131 Trieste, Italy\\
\textsuperscript{22} INAF-Osservatorio Astronomico di Trieste, via G. B. Tiepolo 11, I-34143 Trieste, Italy\\
\textsuperscript{23} Institute for Fundamental Physics of the Universe, Via Beirut 2, 34014 Trieste, Italy\\
\textsuperscript{24} Hamburger Sternwarte, Universit\"\\
\textsuperscript{25} School of Mathematics and Physics, University of Queensland,  Brisbane, QLD 4072, Australia\\
\textsuperscript{26} Department of Physics, IIT Hyderabad, Kandi, Telangana 502285, India\\
\textsuperscript{27} Fermi National Accelerator Laboratory, P. O. Box 500, Batavia, IL 60510, USA\\
\textsuperscript{28} Department of Physics and Astronomy, University of Waterloo, 200 University Ave W, Waterloo, ON N2L 3G1, Canada\\
\textsuperscript{29} Department of Physics, University of Michigan, Ann Arbor, MI 48109, USA\\
\textsuperscript{30} California Institute of Technology, 1200 East California Blvd, MC 249-17, Pasadena, CA 91125, USA\\
\textsuperscript{31} Departments of Statistics and Data Sciences, University of Texas at Austin, Austin, TX 78757, USA\\
\textsuperscript{32} NSF-Simons AI Institute for Cosmic Origins, University of Texas at Austin, Austin, TX 78757, USA\\
\textsuperscript{33} SLAC National Accelerator Laboratory, Menlo Park, CA 94025, USA\\
\textsuperscript{34} Instituto de Fisica Teorica UAM/CSIC, Universidad Autonoma de Madrid, 28049 Madrid, Spain\\
\textsuperscript{35} Universität Innsbruck, Institut für Astro- und Teilchenphysik, Technikerstr. 25/8, 6020 Innsbruck, Austria\\
\textsuperscript{36} Center for Astrophysical Surveys, National Center for Supercomputing Applications, 1205 West Clark St., Urbana, IL 61801, USA\\
\textsuperscript{37} Department of Astronomy, University of Illinois at Urbana-Champaign, 1002 W. Green Street, Urbana, IL 61801, USA\\
\textsuperscript{38} School of Physics and Astronomy, Cardiff University, CF24 3AA, UK\\
\textsuperscript{39} Santa Cruz Institute for Particle Physics, Santa Cruz, CA 95064, USA\\
\textsuperscript{40} Center for Cosmology and Astro-Particle Physics, The Ohio State University, Columbus, OH 43210, USA\\
\textsuperscript{41} Department of Physics, The Ohio State University, Columbus, OH 43210, USA\\
\textsuperscript{42} Department of Astronomy/Steward Observatory, University of Arizona, 933 North Cherry Avenue, Tucson, AZ 85721-0065, USA\\
\textsuperscript{43} Jet Propulsion Laboratory, California Institute of Technology, 4800 Oak Grove Dr., Pasadena, CA 91109, USA\\
\textsuperscript{44} Centre for Gravitational Astrophysics, College of Science, The Australian National University, ACT 2601, Australia\\
\textsuperscript{45} The Research School of Astronomy and Astrophysics, Australian National University, ACT 2601, Australia\\
\textsuperscript{46} Departamento de F\'isica Matem\'atica, Instituto de F\'isica, Universidade de S\~ao Paulo, CP 66318, S\~ao Paulo, SP, 05314-970, Brazil\\
\textsuperscript{47} Faculty of Physics, Ludwig-Maximilians-Universit\"at, Scheinerstr. 1, 81679 Munich, Germany\\
\textsuperscript{48} Max Planck Institute for Extraterrestrial Physics, Giessenbachstrasse, 85748 Garching, Germany\\
\textsuperscript{49} George P. and Cynthia Woods Mitchell Institute for Fundamental Physics and Astronomy, and Department of Physics and Astronomy, Texas A\&M University, College Station, TX 77843,  USA\\
\textsuperscript{50} Department of Astrophysical Sciences, Princeton University, Peyton Hall, Princeton, NJ 08544, USA\\
\textsuperscript{51} Kavli Institute for Particle Astrophysics \& Cosmology, P. O. Box 2450, Stanford University, Stanford, CA 94305, USA\\
\textsuperscript{52} LPSC Grenoble - 53, Avenue des Martyrs 38026 Grenoble, France\\
\textsuperscript{53} Instituci\'o Catalana de Recerca i Estudis Avan\c\\
\textsuperscript{54} Department of Physics, University of Cincinnati, Cincinnati, Ohio 45221, USA\\
\textsuperscript{55} Perimeter Institute for Theoretical Physics, 31 Caroline St. North, Waterloo, ON N2L 2Y5, Canada\\
\textsuperscript{56} Observat\'orio Nacional, Rua Gal. Jos\'e Cristino 77, Rio de Janeiro, RJ - 20921-400, Brazil\\
\textsuperscript{57} Department of Physics, Carnegie Mellon University, Pittsburgh, Pennsylvania 15312, USA\\
\textsuperscript{58} Ruhr University Bochum, Faculty of Physics and Astronomy, Astronomical Institute, German Centre for Cosmological Lensing, 44780 Bochum, Germany\\
\textsuperscript{59} Nordita, KTH Royal Institute of Technology and Stockholm University, Hannes Alfv\'ens v\"ag 12, SE-10691 Stockholm, Sweden\\
\textsuperscript{60} Department of Physics, University of Arizona, Tucson, AZ 85721, USA\\
\textsuperscript{61} Physics Department, Lancaster University, Lancaster, LA1 4YB, UK\\
\textsuperscript{62} Computer Science and Mathematics Division, Oak Ridge National Laboratory, Oak Ridge, TN 37831\\
\textsuperscript{63} Department of Physics, Duke University Durham, NC 27708, USA\\
\textsuperscript{64} Department of Astronomy, Ohio State University, Columbus, OH 43210, USA\\
\textsuperscript{65} Department of Astronomy, University of California, Berkeley,  501 Campbell Hall, Berkeley, CA 94720, USA\\
\textsuperscript{66} Lawrence Berkeley National Laboratory, 1 Cyclotron Road, Berkeley, CA 94720, USA\\
\textsuperscript{67} Department of Physics, Stanford University, 382 Via Pueblo Mall, Stanford, CA 94305, USA\\
\textsuperscript{68} Universit\"ats-Sternwarte, Fakult\"at f\"ur Physik, Ludwig-Maximilians Universit\"at M\"unchen, Scheinerstr. 1, 81679 M\"unchen, Germany\\
\textsuperscript{69} Department of Physics, Southern Methodist University, Dallas, TX 75205, USA

\end{document}